\documentclass[11pt,a4paper]{article}
\usepackage{amssymb,amsmath,mathrsfs}
\usepackage{fullpage}
\usepackage{natbib}
\usepackage{hyperref}
\usepackage{upgreek}
\usepackage{lineno}
\usepackage{graphicx}

\def\d{\mathrm{d}}

\def\beq{\begin{equation}}
\def\eeq{\end{equation}}

\def\bx{\boldsymbol{x}}

\def\bu{\boldsymbol{u}}

\def\ip{\lrcorner}

\newcommand{\dt}[2]{\frac{\mathrm{d} #1}{\mathrm{d} #2}}

\newcommand{\eqn}[1]{(\ref{eqn:#1})}
\newcommand{\lab}[1]{\label{eqn:#1}}
\newcommand{\inter}[1]{\quad \textrm{#1} \quad}

\newcommand{\bs}[1]{\boldsymbol{#1}}

\def\lie{\mathcal{L}}

\def\Laplace{\Updelta}

\newcommand{\divv}{\mathop{\mathrm{div}}\nolimits}

\def\XXint#1#2#3{{\setbox0=\hbox{$#1{#2#3}{\int}$}
\vcenter{\hbox{$#2#3$}}\kern-.5\wd0}}

\newcommand*\patchAmsMathEnvironmentForLineno[1]{%
  \expandafter\let\csname old#1\expandafter\endcsname\csname #1\endcsname
  \expandafter\let\csname oldend#1\expandafter\endcsname\csname end#1\endcsname
  \renewenvironment{#1}%
     {\linenomath\csname old#1\endcsname}%
     {\csname oldend#1\endcsname\endlinenomath}}%
\newcommand*\patchBothAmsMathEnvironmentsForLineno[1]{%
  \patchAmsMathEnvironmentForLineno{#1}%
  \patchAmsMathEnvironmentForLineno{#1*}}%
\AtBeginDocument{%
\patchBothAmsMathEnvironmentsForLineno{equation}%
\patchBothAmsMathEnvironmentsForLineno{align}%
\patchBothAmsMathEnvironmentsForLineno{flalign}%
\patchBothAmsMathEnvironmentsForLineno{alignat}%
\patchBothAmsMathEnvironmentsForLineno{gather}%
\patchBothAmsMathEnvironmentsForLineno{multline}%
}

\title{Vortex dynamics on a M\"obius strip}
\author{Jacques Vanneste}

\date{\small School  of  Mathematics  and  Maxwell  Institute  for  Mathematical  Sciences,  University  of  Edinburgh, King’s Buildings, Edinburgh EH9 3FD, United Kingdom}

\begin{document}

\maketitle

\begin{abstract}
\noindent
We consider the dynamics of a two-dimensional incompressible perfect fluid on a M\"obius strip embedded in $\mathbb{R}^3$. The vorticity--streamfunction formulation of the Euler equations is derived from an exterior-calculus form of the momentum equation. The non-orientability of the M\"obius strip and the distinction between forms and pseudo-forms this introduces lead to unusual properties: a boundary condition is provided by the conservation of circulation along the single boundary of the strip, and there is no integral conservation for the vorticity or for any odd function thereof. A finite-difference numerical implementation is used to illustrate the M\"obius-strip realisation of familiar phenomena: translation of vortices along boundaries, shear instability, and decaying turbulence.  
\end{abstract}

\section{Introduction}

In this paper, we examine the dynamics of a two-dimensional incompressible inviscid fluid confined to a M\"obius strip. The main motivation is curiosity about the impact of the non-orientability of the M\"obius strip on the phenomenology of vortex dynamics. A secondary, more frivolous motivation is the wish to produce interesting fluid-dynamical pictures. The system is physically realisable: soap films, for example, can be used as proxys for two-dimensional fluids \citep[e.g.][]{coud-basd,coud-et-al}; with a suitable wire frame they take the topology of a M\"obius strip \citep[e.g.][]{cour40,gold-et-al}. The experiments in this paper are however exclusively numerical.

\begin{figure}
\begin{center}
\includegraphics[height=5.5cm]{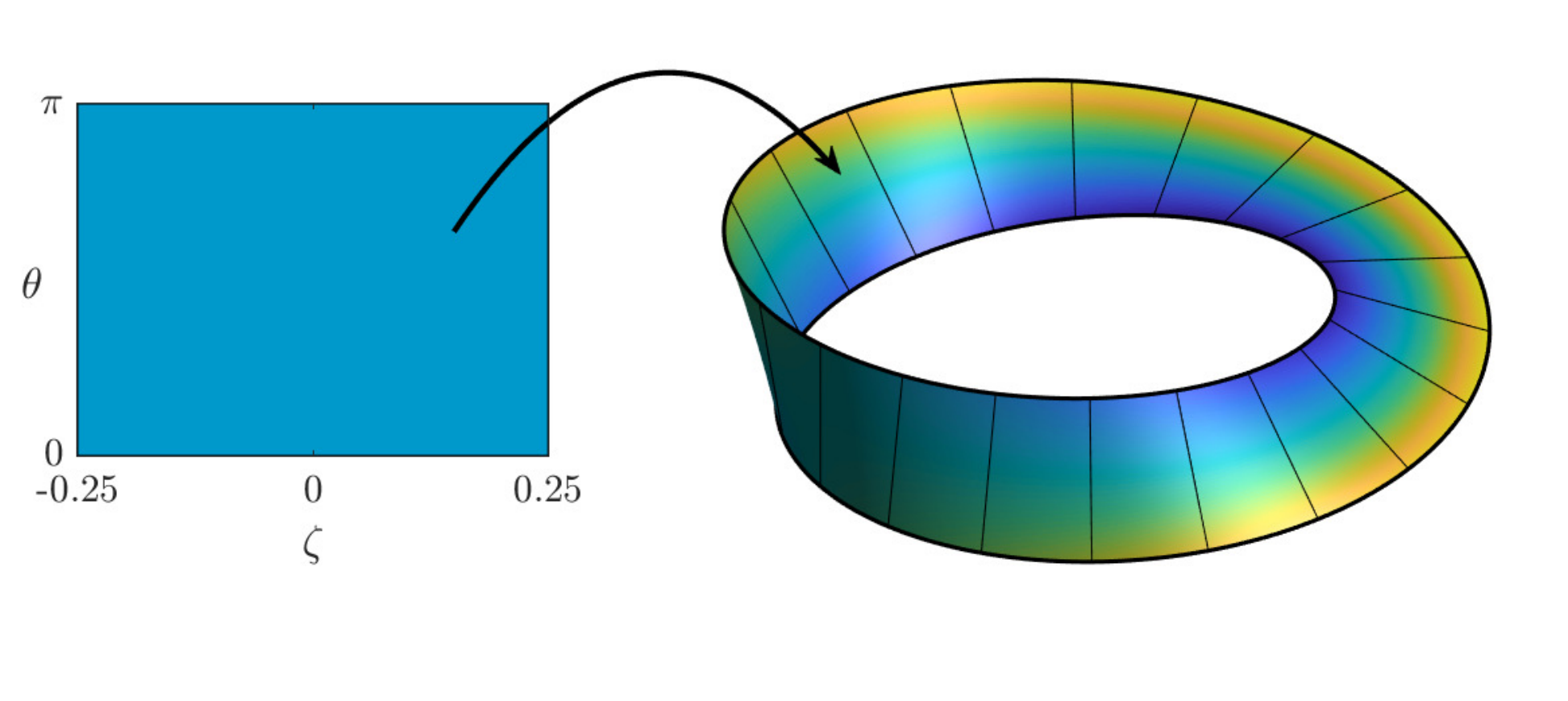}
\caption{M\"obius strip in $\mathbb{R}^3$ and its parameter space for $a=0.25$. The M\"obius strip is coloured according to the value of $|g|^{1/2}$, with $|g|$ the metric determinant in \eqn{|g|}, which ranges from about $1.5$ (dark blue) to $2.5$ (light yellow) and whose variations are associated with curvature effects.}
\label{fig:mobius}
\end{center}
\end{figure}

The M\"obius strip is a topological object but fluid dynamics requires a geometry, which we need to choose. The simplest geometry is that of a flat strip, imposing antisymmetry conditions across a `seam' to achieve the necessary twist. However, the flat M\"obius strip cannot be embedded in $\mathbb{R}^3$, a necessary condition for  physical realisability. We therefore have chosen another widely used model, the ruled surface constructed by rotating the midpoint of a segment along a horizontal circle while rotating the segment in the vertical at half the rate. This surface, which we denote by $\mathcal{M}$, is parameterised as
\begin{subequations} \lab{mobius}
\begin{align}
x&=(1+\zeta \cos \theta) \cos(2 \theta), \\
y&=(1+\zeta \cos \theta) \sin(2 \theta), \\
z&=\zeta \sin \theta,
\end{align}
\end{subequations}
with $-a \le \zeta \le a$ and $0 \le \theta \le \pi$ (or rather $\theta$ interpreted as taken modulo $\pi$). The property {$\bx(\zeta,\theta+\pi)=\bx(-\zeta,\theta)$} reflects the twist of the M\"obius-strip topology. The parameterisation and shape of $\mathcal{M}$ are illustrated in Fig.\ \ref{fig:mobius}. The mean curvature of $\mathcal{M}$ does not vanish, so it is not a minimal surface as formed by a soap film. Its Gaussian curvature does not vanish either, so it is not developable and cannot be made of an inextensible material. Closed-form expressions are available for M\"obius strips in $\mathbb{R}^3$ that are either minimal \citep{odeh16} or developable \citep[see][]{schw90}
and could replace \eqn{mobius} to gain realism. The equilibrium shape taken by an inextensible material twisted as a M\"obius strip can also be obtained, but this requires solving an energy-minimisation problem numerically \citep{star-van}.

The non-orientability of $\mathcal{M}$ is manifested in the fact that the basis  $(\partial_{\zeta} \bx,\partial_\theta \bx)$ of its tangent space has opposite orientations for $\theta=0$ and $\theta=\pi$. Equivalently, if a normal vector $\bs{N}$ is added to $\partial_{\zeta} \bx$ and $\partial_\theta \bx$ to make up a right-handed basis of $\mathbb{R}^3$, it satisfies $\bs{N}(\theta=\pi)=-\bs{N}(\theta=0)$. 
Non-orientability is a global property, not discernible at the level of a local patch of the strip; incompressible fluid dynamics is also global, through the pressure field. Nonetheless, for  narrow strips such as the one in Fig.\ \ref{fig:mobius}, and with the scale of fluid structures set by the strip's width, we can expect the non-orientability to have a limited impact on the dynamics, which differs from that in a flat cylinder (channel) primarily because of the strip's curvature. The main effect of non-orientability arises from the fact that the sense of rotation of vortices is reversed when they are transported once along the entire length of the strip, that is, when $\theta$ increases by $\pi$. We will illustrate this by considering the dynamics of a single vortex propagating along the strip's boundary. A mathematical consequence is that  vorticity cannot be meaningfully integrated over the M\"obius strip; as a result, the familiar connection between circulation and vorticity integral is lost, and the conservation of integrals of functions of the vorticity (Casimirs) is restricted to even functions.

The paper is structured as follows. In \S\ref{sec:model}, we derive the vorticity--streamfunction formulation of the Euler equations for an incompressible perfect fluid on $\mathcal{M}$ in terms of the coordinates $(\zeta,\theta)$. We start with a convenient formulation of the Euler equations that highlights the dual role of the vector and 1-form fields associated with velocity, and treats  vorticity as a 2-form. We carefully distinguish between forms (such as the vorticity 2-form) and pseudo-forms (such as the area 2-form, the streamfunction and a pseudo-scalar version of the vorticity {which we term `vorticity density'}). The distinction is crucial for a non-orientable surface such as $\mathcal{M}$. We derive the boundary conditions for the vorticity--streamfunction formulation; these include a condition involving the circulation along the (single) boundary of $\mathcal{M}$. In \S\ref{sec:numerics} we describe a numerical implementation of the equations which we employ to examine the propagation of a vortex along the strip's boundary, shear instability and decaying turbulence. Section \ref{sec:final} summarises the main features of vortex dynamics on a non-orientable surface. Two appendices provide mathematical details.

\section{Euler equations on the M\"obius strip} \label{sec:model}

A coordinate expression for the Euler equations governing the dynamics of an incompressible perfect fluid on an arbitrary surface was derived by \citet{zerm02} (see \citet{zerm04} for a commented English translation). Modern formulations with a focus on point vortices and closed surfaces can be found in, e.g.,\  \citet{kimu99}, \citet{boat-koil}, \citet{drit-boat}, \citet{raga-barr} or \citet{gust19}.
While Zermelo's expressions can be readily particularised to the M\"obius strip, it is instructive to start from a general, coordinate-free version of the Euler equations. A convenient form is
\begin{subequations} \lab{euler}
\begin{align}
(\partial_t + \lie_{\bu}) \nu &= - d \left( p - \tfrac{1}{2} | \bs{u}|^2 \right), \lab{momentum} \\ \quad \divv \bu & = 0. \lab{incomp}
\end{align}
\end{subequations}
Here $\bu$ is the velocity (vector) field, $\nu$ is the momentum 1-form, obtained from $\bu$ and the metric $g$ as $\nu = g(\bu,\cdot)$,  $\lie_{\bu}$ is the Lie derivative  and $d$ the exterior derivative \citep[see e.g.][]{schu80,arno-kesh,bess-fris,gilb-v18}. 

The metric on the M\"obius strip $\mathcal{M}$ is found by pulling back the Euclidean metric $dx \otimes dx + dy \otimes dy + dz \otimes dz$ of $\mathbb{R}^3$, that is, by using \eqn{mobius} to express $dx$, $dy$ and $dz$ in terms of $d \zeta$ and $d \theta$ to find 
\beq
g = d \zeta \otimes d \zeta + \left(4 + 8 \cos \theta \, \zeta + (3+2 \cos(2 \theta)) \zeta^2\right) d \theta\otimes d \theta.
\lab{metric2}
\eeq
This is a diagonal tensor, indicating that the coordinates $(\zeta,\theta)$ are orthogonal {(i.e.\, the basis vectors $\partial_\zeta$ and $\partial_\theta$ are orthogonal)} with determinant
\beq
|g| = 4 + 8 \cos \theta \, \zeta + \left(3+2 \cos(2 \theta)\right) \zeta^2.
\lab{|g|}
\eeq
{It follows from \eqn{metric2} that the vectors $\partial_\zeta$ and $\partial_\theta$ have norms $|\partial_\zeta| =1$ and $|\partial_\theta|=|g|^{1/2}$.}
Writing the velocity field in components as
\beq
\bu = u^\zeta \partial_\zeta + u^\theta \partial_\theta,
\lab{velcomp}
\eeq
the corresponding momentum 1-form is found from $\nu=g(\bu,\cdot)$ as
\beq
\nu = \nu_\zeta \, d \zeta + \nu_\theta  \, d \theta = u^\zeta \, d \zeta + |g| u^\theta \, d \theta.
\lab{nucomp}
\eeq
Introducing \eqn{nucomp} into \eqn{momentum} with $\lie_{\bu} =  u^\zeta \partial_\zeta + u^\theta \partial_\theta$ and using the commutation $\lie_{\bu} d = d \lie_{\bu}$ readily gives the momentum equation in terms of $u^\zeta$ and $u^\theta$. This is worked out in Appendix \ref{app:momentum}.

The divergence in \eqn{incomp} is defined by the relation $\divv \bu \, \mu = d (\bu \ip \mu) = \lie_{\bu} \mu $, where $\mu$ is the volume form associated with the metric $g$ (area form since $\mathcal{M}$ is two-dimensional) and $\ip$ denotes the interior product or contraction. The area form  is given by
\beq
\mu = \pm |g|^{1/2} d \zeta \wedge d \theta.
\lab{mu2}
\eeq
This is a \textit{pseudo}-2-form, which changes sign when the orientation of the basis $(\partial_\zeta \bx,\partial_\theta \bx)$ changes, as indicated by the notation $\pm$. Since $\mathcal{M}$ is non-orientable, a change in the orientation of the basis is unavoidable; it occurs across $\theta= 0 \mod \pi$ in the parameterisation \eqn{mobius}. It can be verified that \eqn{mu2} is the projection of the Euclidean volume $d x \wedge dy \wedge d z$ on the normal $\bs{N}=(N^1,N^2,N^3)$ to the M\"obius strip, that is,  
$\mu = {\bs{N}} \ip \, dx \wedge d y \wedge dz = N^1 dy \wedge dz + N^2 dz \wedge \d x + N^3 \d x \wedge dy$. With \eqn{mu2} we compute
\beq
\bu \ip \mu =  \pm |g|^{1/2} (u^\zeta \, d \theta - u^\theta \, d \zeta),
\lab{buipmu}
\eeq
and apply $d$ to obtain the incompressibility condition in the explicit form
\beq
\divv \bu = {|g|^{-1/2}} \left(\partial_\zeta ( |g|^{1/2} u^\zeta) + \partial_\theta (|g|^{1/2} u^\theta) \right) = 0.
\lab{incomp3}
\eeq

The momentum and incompressibility equations \eqn{euler3} and \eqn{incomp3} can be conveniently replaced by the vorticity equation. This is best derived directly from \eqn{momentum}: applying $d$ and using that $d$ and $\lie_{\bu}$ commute and that $d^2=0$ gives
\beq
(\partial_t + \lie_{\bu}) d \nu = 0,
\lab{vorticity}
\eeq 
showing that the vorticity 2-form $d \nu$ is transported by the flow. A vorticity density $\omega$ can be defined by
\beq
\omega \mu = d \nu.
\lab{omega}
\eeq
This is a \textit{pseudo}-scalar, which, like $\mu$, changes sign with the orientation of the basis on $\mathcal{M}$. Applying $d$ to \eqn{velcomp} and using \eqn{mu2} gives the explicit expression
\beq
\omega = \pm |g|^{-1/2} \left(\partial_\zeta (|g| u^\theta) - \partial_\theta u^\zeta \right).
\lab{omega3}
\eeq
Incompressibility ensures that $(\partial_t + \lie_{\bu}) \mu = 0$, hence $\omega$ is also transported by the flow:
\beq
(\partial_t + \lie_{\bu}) \omega =0.
\lab{vort1}
\eeq
The velocity $\bu$ can be related to the vorticity by means of a streamfunction $\psi$, with 
\beq
u^\zeta = - |g|^{-1/2} \partial_\theta \psi \inter{and}
u^\theta = |g|^{-1/2} \partial_\zeta \psi,
\lab{upsi}
\eeq 
so as to satisfy the incompressibility condition \eqn{incomp3}. Equivalently, this can be written as
\beq
d \psi = - \bu \ip \mu,
\lab{psiu}
\eeq
where $\ip$ denotes the interior product.
With \eqn{upsi}, \eqn{vort1} becomes
\beq
\partial_t \omega + |g|^{-1/2} \partial(\psi,\omega)=0,
\lab{vort2}
\eeq
where $\partial(\psi,\omega) = \partial_\zeta \psi \partial_\theta \omega - \partial_\theta \psi \partial_\zeta \omega$ is the Jacobian operator. Substituting \eqn{upsi} into \eqn{omega3} then gives the vorticity--streamfunction relation
%
\beq
\omega = \Laplace \psi = {|g|^{-1/2}} \left(\partial_\zeta \left(|g|^{1/2} \partial_\zeta \psi\right) + \partial_\theta \left(|g|^{-1/2} \partial_\theta \psi \right) \right),
\lab{laplace}
\eeq
where we have identified the Laplacian $\Laplace$. A completely geometric derivation of \eqn{upsi} and \eqn{laplace}, which avoids coordinates but requires the introduction of the Hodge * operator, is given in Appendix \ref{app:hodge}.

Eqs.\ \eqn{vort2}--\eqn{laplace}, together with the form \eqn{|g|} of the metric determinant $|g|$ constitute the vorticity--streamfunction formulation of the Euler equations on the M\"obius strip. In the familiar way, the vorticity {density} $\omega$ is the dynamical variable from which the streamfunction $\psi$ is derived by inverting the Poisson equation \eqn{laplace}. This requires boundary conditions. Because $\omega$ and $\psi$ are pseudo-scalars, they satisfy
\beq
\omega(\zeta,\theta+\pi,t) = -\omega(-\zeta,\theta,t) \inter{and}
\psi(\zeta,\theta+\pi,t) = -\psi(-\zeta,\theta,t).
\lab{pseudo-scalar}
\eeq
In particular, 
\beq
\psi(\zeta,\pi,t)=-\psi(-\zeta,0,t),
\lab{bc1}
\eeq
which provides a first boundary condition for \eqn{laplace}.
{The no-normal-flow condition 
\beq
u^\zeta(\zeta=\pm a,\theta,t)=0
\eeq
and \eqn{upsi} imply} that $\psi$ depends only on $t$ along the boundary of the strip. Because of the pseudo-scalar nature of $\psi$, this constancy translates into
\beq
\psi (\zeta = \pm a, \theta,t)= \pm C(t).
\lab{bc2}
\eeq 
This provides the second boundary condition for the Poisson equation \eqn{laplace}. To determine $C(t)$, we need to revert to the momentum formulation \eqn{euler}. It implies conservation of the circulation along the (single) boundary, given by
\begin{align}
\Gamma &= \int_{\partial \mathcal{M}} \nu = \int_0^\pi \left(\nu_\theta(a,\theta,t)+\nu_\theta(-a,\theta,t)\right) \, d \theta \nonumber \\
&= \int_0^\pi \left({|g(a,\theta)|^{1/2}} \partial_\zeta \psi (a,\theta,t)+{|g(-a,\theta)|^{1/2}} \partial_\zeta \psi (-a,\theta,t)\right) \, d \theta.
\lab{circ}
\end{align}
The equation 
\beq
\dt{\Gamma}{t}=0
\lab{circcons}
\eeq
completes the formulation. The description of our numerical implementation in \S\ref{sec:numerics} makes it clear how \eqn{circcons} determines  $C(t)$ for use in \eqn{bc2}.

The existence of infinitely many conserved integrals is a key feature of two-dimensional inviscid incompressible fluids. Here the non-orientablity of the M\"obius strip leads to important differences compared with the familiar, orientable case. This is because, while pseudo-2-forms can be integrated over non-orientable surfaces, proper 2-forms cannot -- the result of such an integration would not be coordinate independent \citep[e.g.][]{fran04}. Thus while the area of the M\"obius strip $\int_\mathcal{M} \mu$ is well defined, the integral $\int_\mathcal{M} d \nu$ of the vorticity {2-form} is not (and cannot be related to the circulation $\Gamma$). As a result, conserved  Casimirs functions 
\beq
\mathcal{C}_f =  \int_{\mathcal{M}} f(\omega) \mu =  \int_{-a}^a \int_0^\pi  f(\omega) |g|^{1/2} \, d \zeta d \theta,
\eeq 
can only be defined for arbitrary \textit{even} functions $f$. The evenness of $f$ is required to ensure that the change of sign of $\omega$ which accompanies a change of basis orientation has no effect on $f(\omega)$, so that $f(\omega) \mu$ is a pseudo-2-form. The energy
\beq
\mathcal{E} = - \frac{1}{2} \int_M \psi \omega \mu = - \frac{1}{2} \int_{-a}^a \int_0^\pi  \psi \omega |g|^{1/2} \, d \zeta d \theta
\eeq
is also well defined and conserved.

\section{Numerical simulations} \label{sec:numerics}

\subsection{Numerical implementation}

We solve the vorticity--streamfunction equations \eqn{vort2}--\eqn{laplace} numerically.
To control the generation of fine scales that inevitably accompanies non-trivial flows, we modify \eqn{vort2}, adding a small-scale  dissipation to obtain
\beq
\partial_t \omega + |g|^{-1/2} \partial(\psi,\omega) = \varepsilon \Laplace \omega,
\lab{vis}
\eeq
with $\varepsilon>0$ small. This requires an additional boundary condition, taken as $\omega=0$ on the strip boundary $\partial \mathcal{M}$, corresponding to a no-stress condition.
With dissipation, the circulation $\Gamma$ along $\partial \mathcal{M}$ changes. We derive the form of $\d \Gamma/\d t$ in Appendix \ref{app:hodge} where it is given as \eqn{circchange}. {Note that the dissipation model chosen is not standard Newtonian friction: the complete Navier--Stokes vorticity equation involves geometric terms omitted from the right-hand side of \eqn{vis} \citep[see Appendix \ref{app:hodge},][]{gilb-et-al,gilb-v21}. In the simulations of \S\ref{sec:simul}, $\varepsilon$ is taken small enough the results are representative of the limit $\varepsilon \to 0$ except for scales close to the grid size.}

\begin{figure}
\begin{center}
\includegraphics[height=5cm]{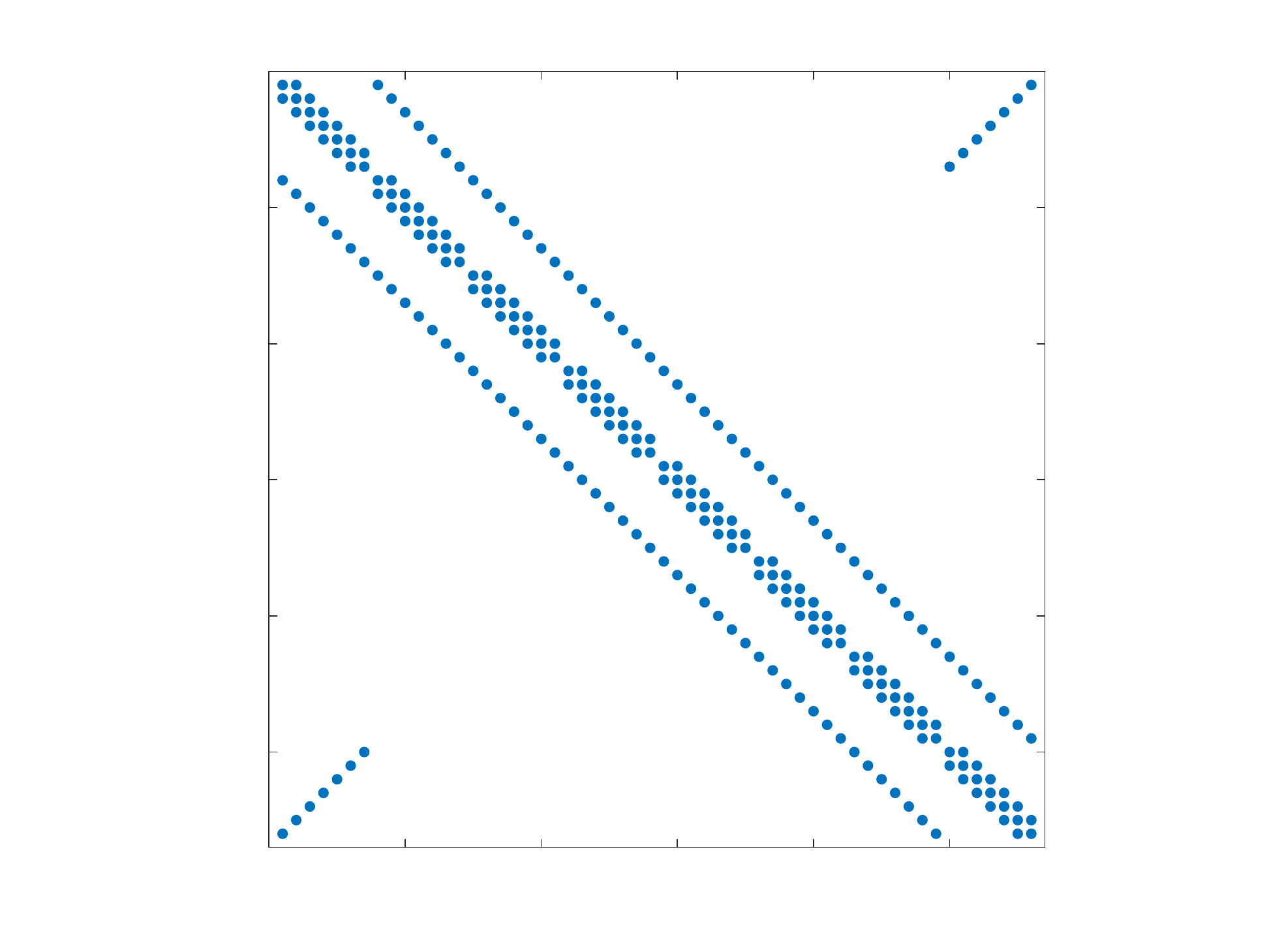}
\caption{Visualisation of the non-zero entries in the matrix discretising the Laplacian operator on the M\"obius strip. The unusual anti-diagonal bands in the top-right and bottom-left corners result from the twist of the M\"obius strip.}
\label{fig:matrix}
\end{center}
\end{figure}

To solve \eqn{laplace} and \eqn{vis} numerically, we discretise $\omega$ and $\psi$ on a regular grid in $(\zeta,\theta)$ coordinates. The Laplacian $\Laplace$ is discretised using a straightforward 5-point stencil, accounting for the condition \eqn{bc1} at the seam $\theta=0$ and assuming homogeneous conditions for $\zeta=\pm a$ corresponding to the strip's boundary $\partial \mathcal{M}$. The resulting matrix is sparse and has an interesting structure displayed in figure \ref{fig:matrix}, with the standard band structure modified by anti-diagonal blocks in the top-right and bottom-left corners. These result from the twist of condition \eqn{bc1}. The boundary condition \eqn{bc2} is implemented as follows. We first compute a numerical approximation to the (time-independent) harmonic function $\psi_\mathrm{har}$ solving
\beq
\Laplace \psi_\mathrm{har} = 0 \quad \textrm{with} \ \ \psi_\mathrm{har}(\zeta=\pm a,\theta)= \pm 1,
\eeq
then, at each time step, solve the Poisson equation \eqn{laplace} with homogeneous boundary condition
\beq
\Laplace \psi_\mathrm{hom} = \omega \quad \textrm{with} \ \ \psi_\mathrm{hom}(\zeta=\pm a,\theta,t)=0.
\eeq
We compute the circulations $\Gamma_\mathrm{har}$ and $\Gamma_\mathrm{hom}(t)$ associated with each solution by direct evaluation of \eqn{circ}, and we obtain the total circulation $\Gamma(t)$ from the differential equation \eqn{circchange}. The solution of the Poisson equation satisfying the boundary condition \eqn{bc2} is then calculated  as
\beq
\psi = \psi_\mathrm{hom} + C(t) \psi_\mathrm{har},
\eeq
with $C(t)$ given by
\beq
C(t) = \left( \Gamma(t) - \Gamma_\mathrm{hom}(t) \right)/\Gamma_\mathrm{har}
\eeq
to ensure that the circulation associated with $\psi$ is $\Gamma(t)$ and \eqn{bc2} holds.

We use a splitting scheme to advance \eqn{vis} in time, with a three-step Adams--Bashforth scheme for the nonlinear advection relying on the \citet{ara66} discretisation of the Jacobian, and a backward Euler scheme for the viscous diffusion. 
For the simulations reported below, the half-width of the M\"obius strip is set to $a=0.25$; the spatial discretisation uses a $200 \times 1000$ grid in $(\zeta,\theta)$; the time step is $0.2$ and the dissipation parameter is $\epsilon = 4 \times 10^{-7}$. With this small value, energy is approximately conserved over the duration of the simulations presented next. 

\subsection{Simulations} \label{sec:simul}

\begin{figure}
\begin{center}
\includegraphics[height=3.6cm]{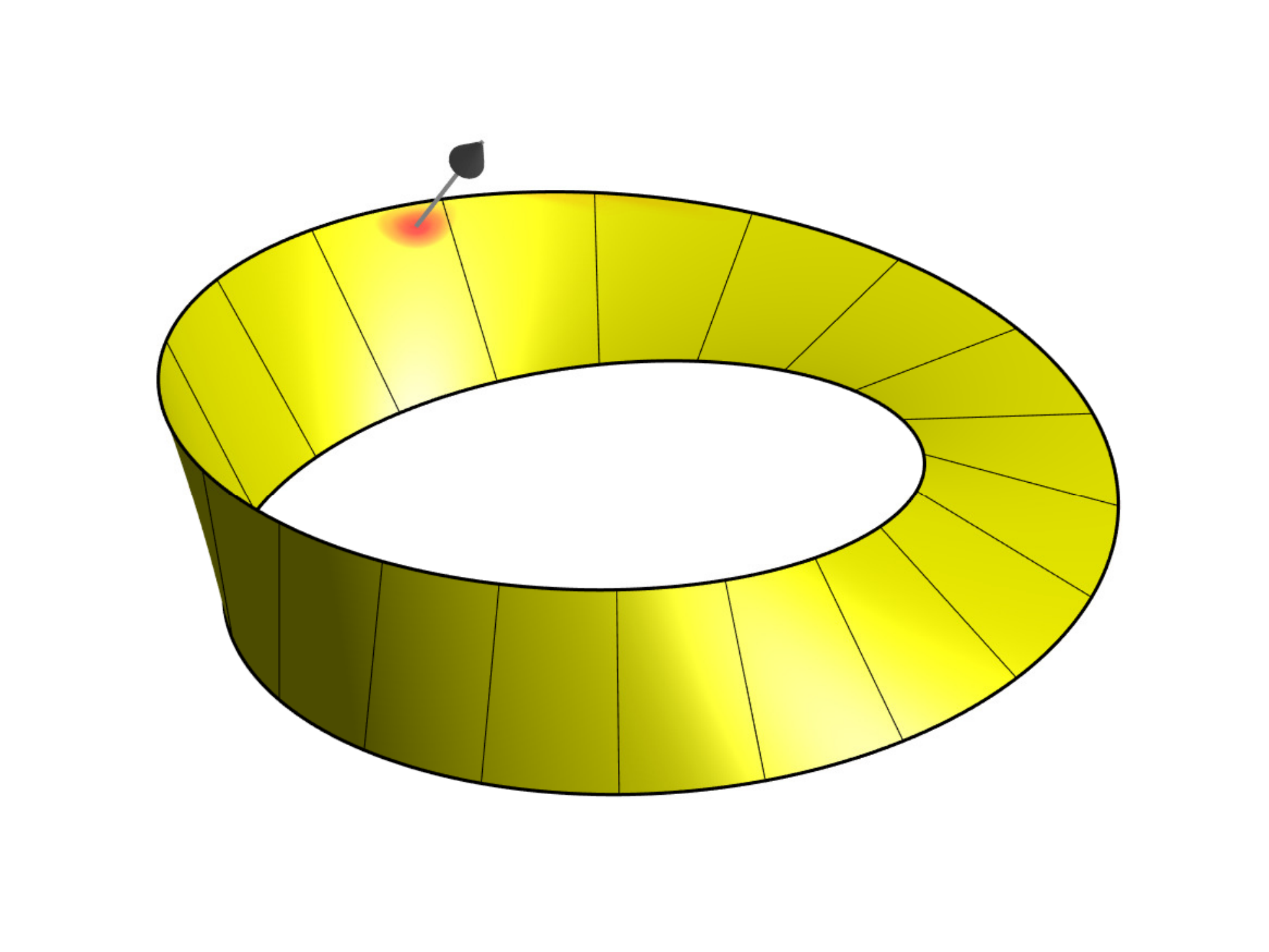} \hspace{-.75cm}
\includegraphics[height=3.6cm]{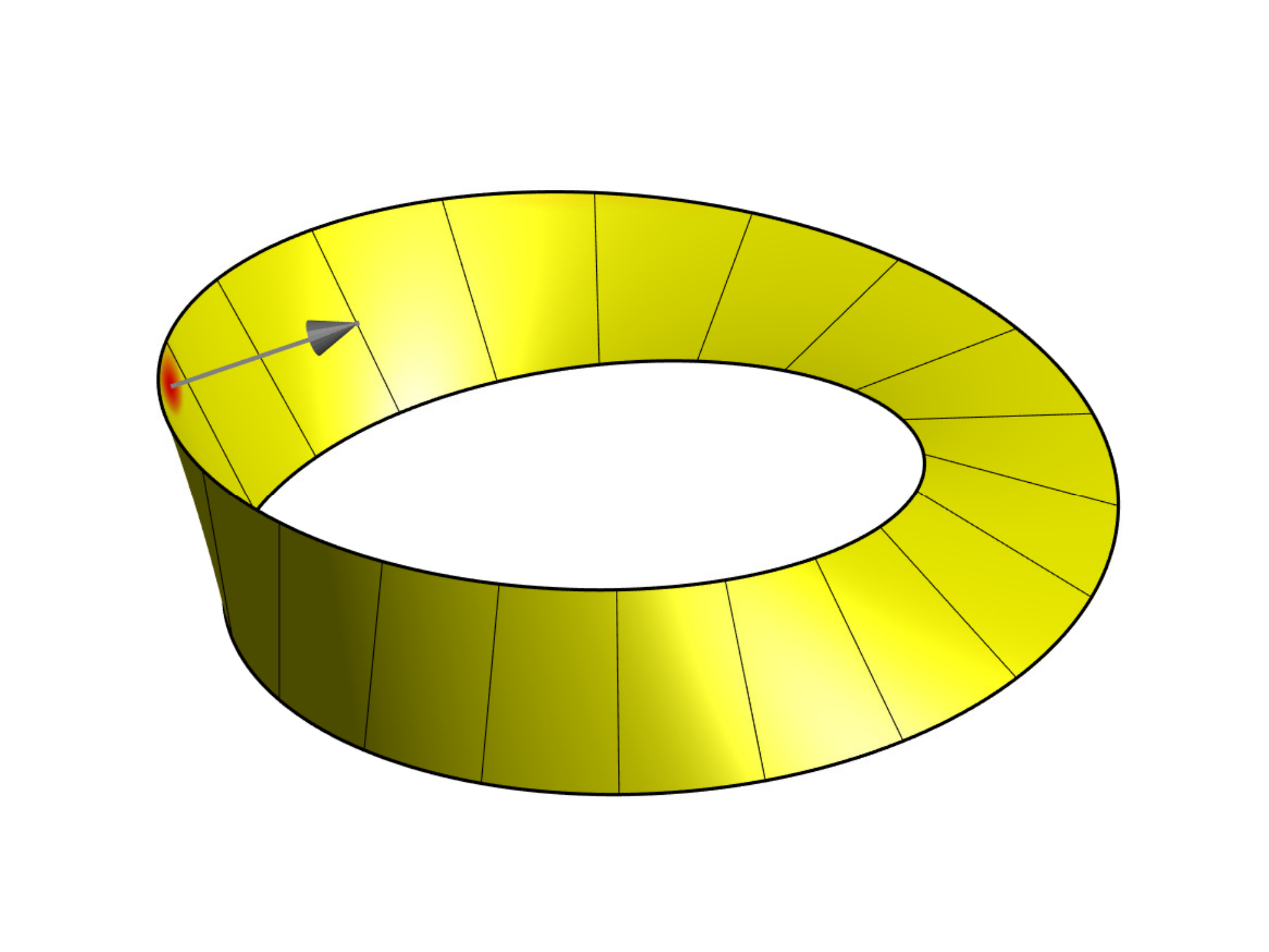} \hspace{-.75cm}
 \includegraphics[height=3.6cm]{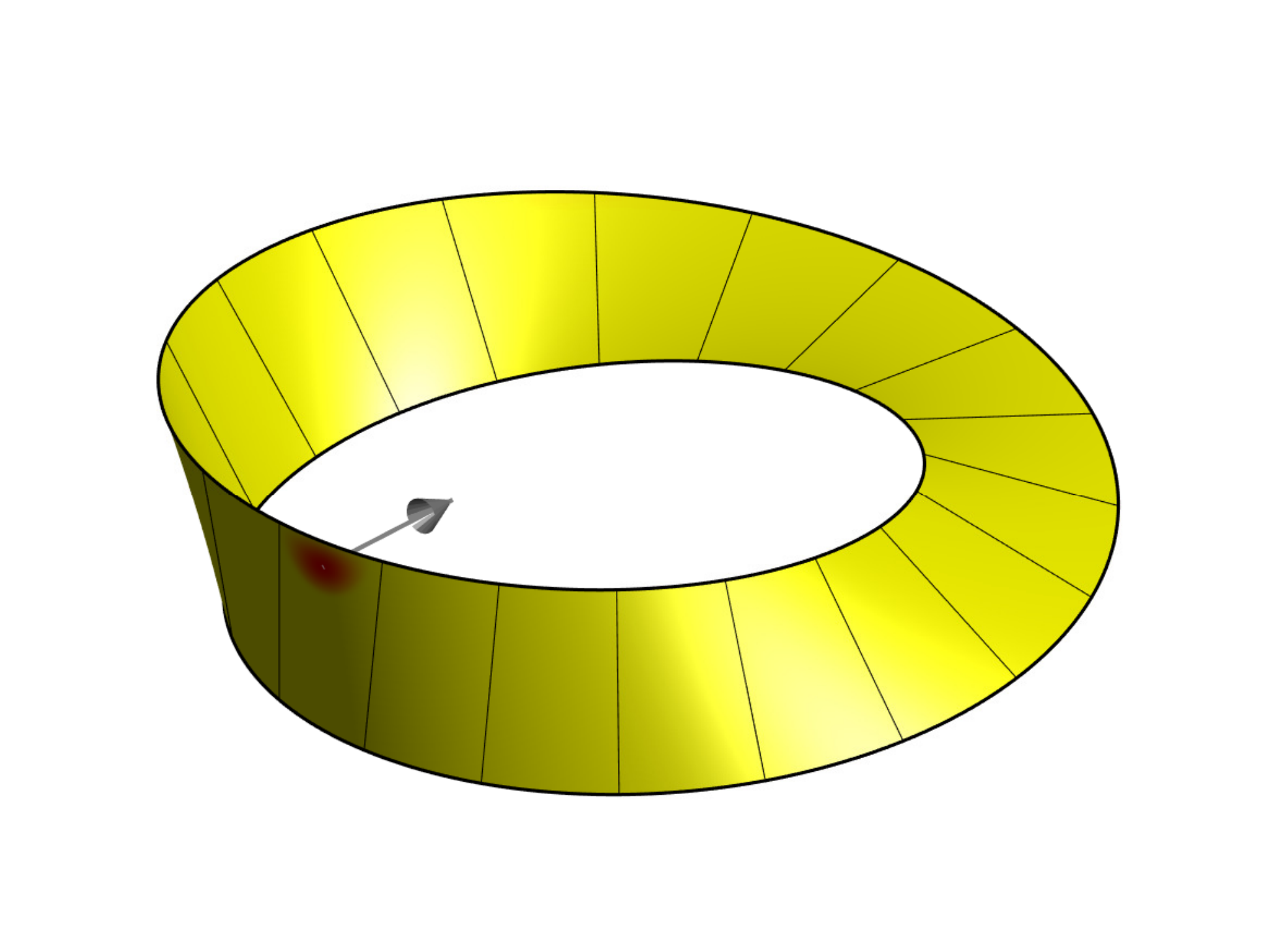}
\vspace{-.5cm} 
 
 \includegraphics[height=3.6cm]{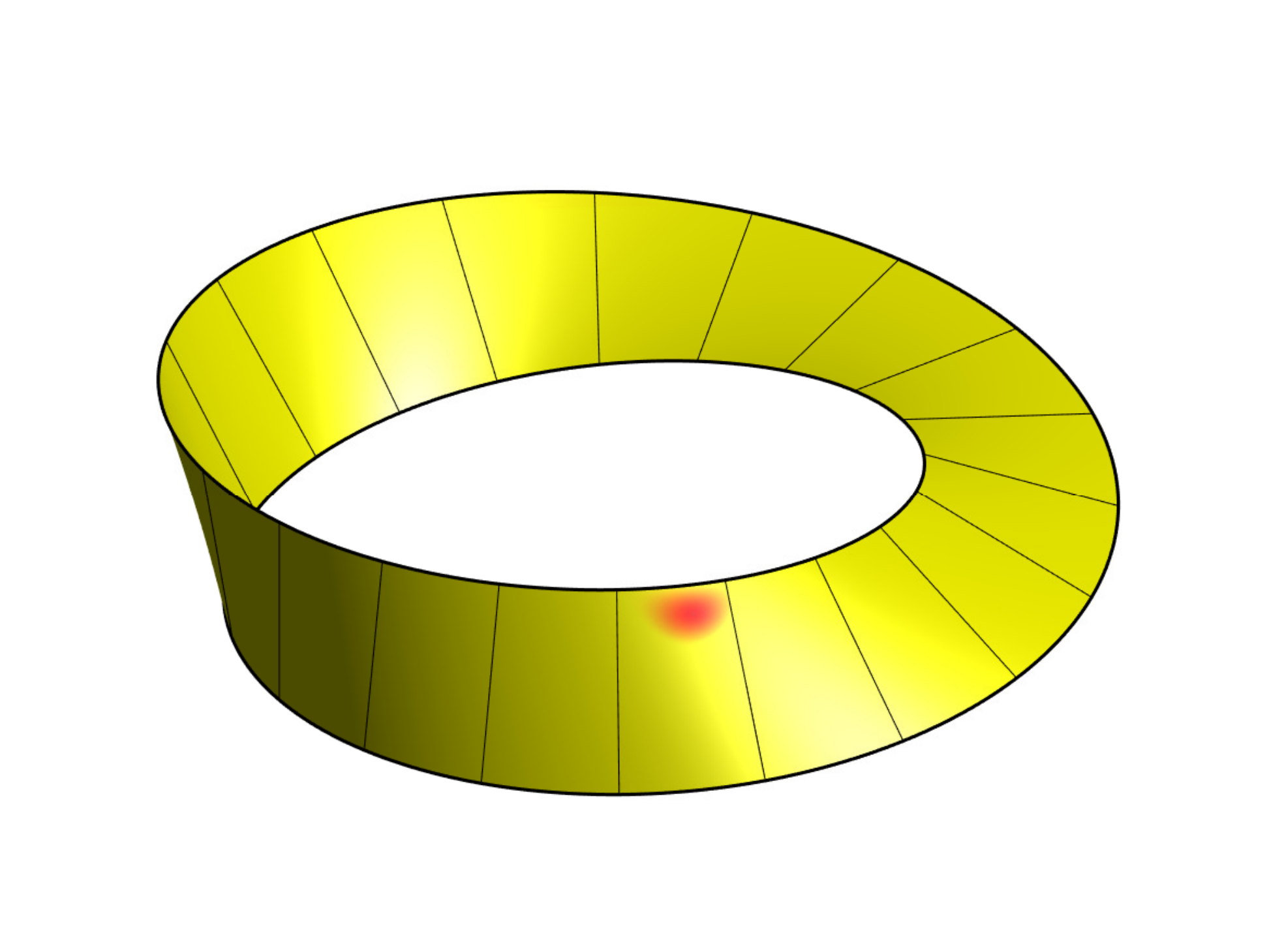} \hspace{-.75cm}
\includegraphics[height=3.6cm]{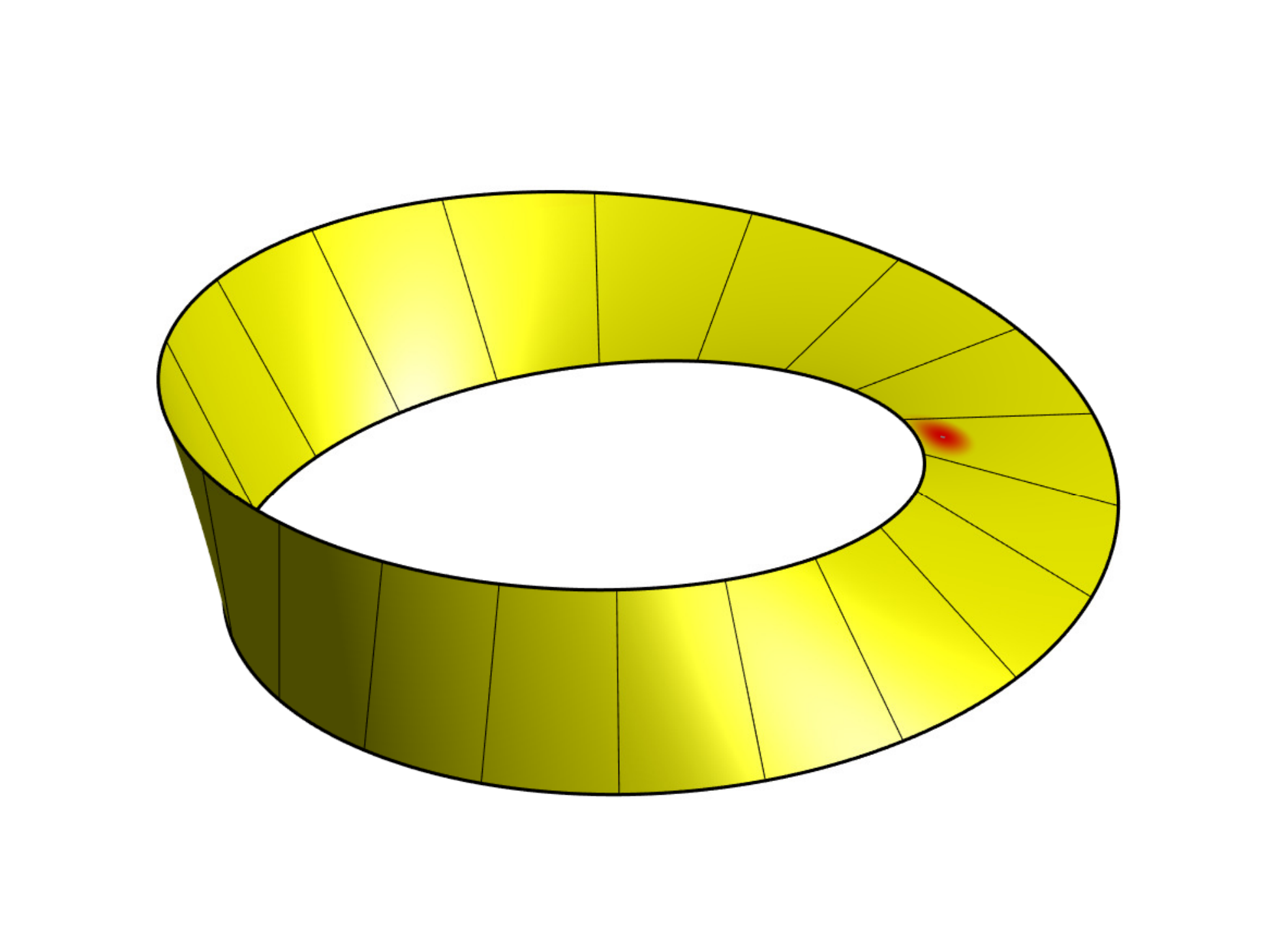} \hspace{-.75cm}
 \includegraphics[height=3.6cm]{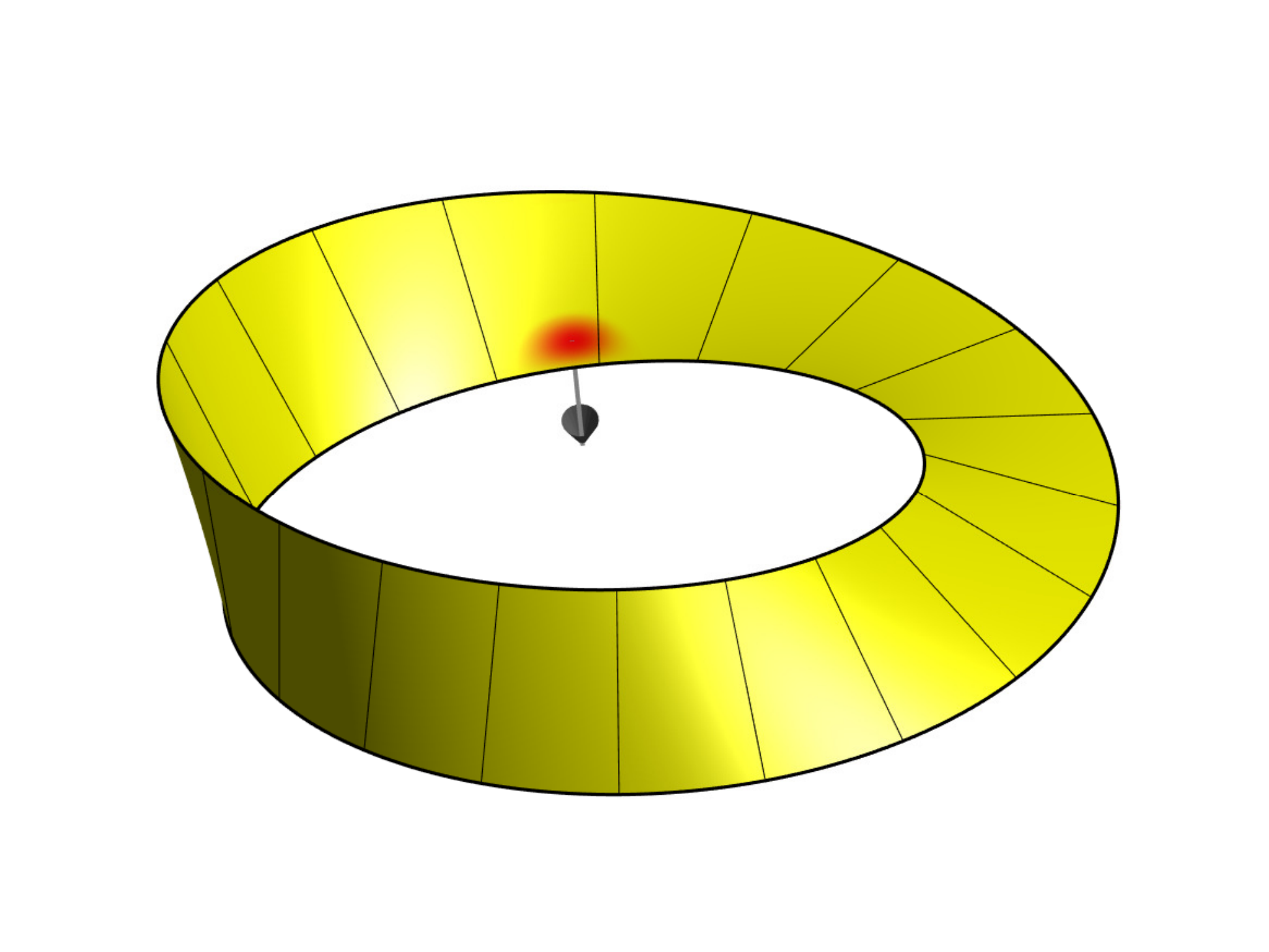} 
 \caption{Propagation of a vortex along the boundary of the M\"obius strip: the vortex travels counterclockwise along the entire boundary, with a vorticity vector field $\omega \bs{N}$ (shown by the arrow in the centre of the vortex) reversing direction as the along-strip coordinate $\theta$ increases by $\pi$ (see also \href{http://www.maths.ed.ac.uk/~vanneste/mobius/vorticityMobiusEdge.mp4}{movie 1}).}
 \label{fig:edge}
\end{center}
\end{figure}

We start by simulating the dynamics of a single vortex patch initialised near the boundary of the M\"obius strip. The vortex is expected to travel along the boundary, forming a dipole with its notional image as is familiar in the planar case. The M\"obius strip case is intriguing in that the strip has a single boundary, {and the direction of rotation of the fluid can be reversed by transporting the vortex once along the strip.} 
The motion of the vortex is in fact straightforward: the vortex travels continuously along the entire boundary. When the along-strip coordinate $\theta$ of the vortex has increased by $\pi$, so that the vortex returns to the same segment $\theta=\mathrm{const}$, the sign of $\omega$ is reversed but the vortex initially near the boundary at $\zeta=a$ is then near $\zeta=-a$ and the propagation direction is unchanged. This is illustrated in the snapshots shown in figure \ref{fig:edge} and (better) in the accompanying \href{http://www.maths.ed.ac.uk/~vanneste/mobius/vorticityMobiusEdge.mp4}{movie 1}. 
These visualise the vorticity 2-form $d \nu$ which, unlike $\omega$, is a true geometric object, independent of the choice of coordinate basis. The field $d \nu$ can be represented as the (true) vector field $\omega \bs{N}$, normal to $\mathcal{M}$ and defined intrinsically as dual to $d\nu$ via the $\mathbb{R}^3$ volume-form $dx \wedge dy \wedge dz$. The visualisation in figure \ref{fig:edge} and movie 1 shows $\omega \bs{N}$ where $|\omega|$ is maximum, that is, at the centre of the vortex, as well as the scalar field $|\omega|$ on the strip. The propagation of a vortex along the edge of the M\"obius strip makes the single-sidedness of the strip manifest and can be viewed as  a fluid-dynamical equivalent of Escher's famous print \textit{M\"obius Strip II} showing ants crawling along the strip. 

\begin{figure}
\begin{center}
\includegraphics[height=3.2cm]{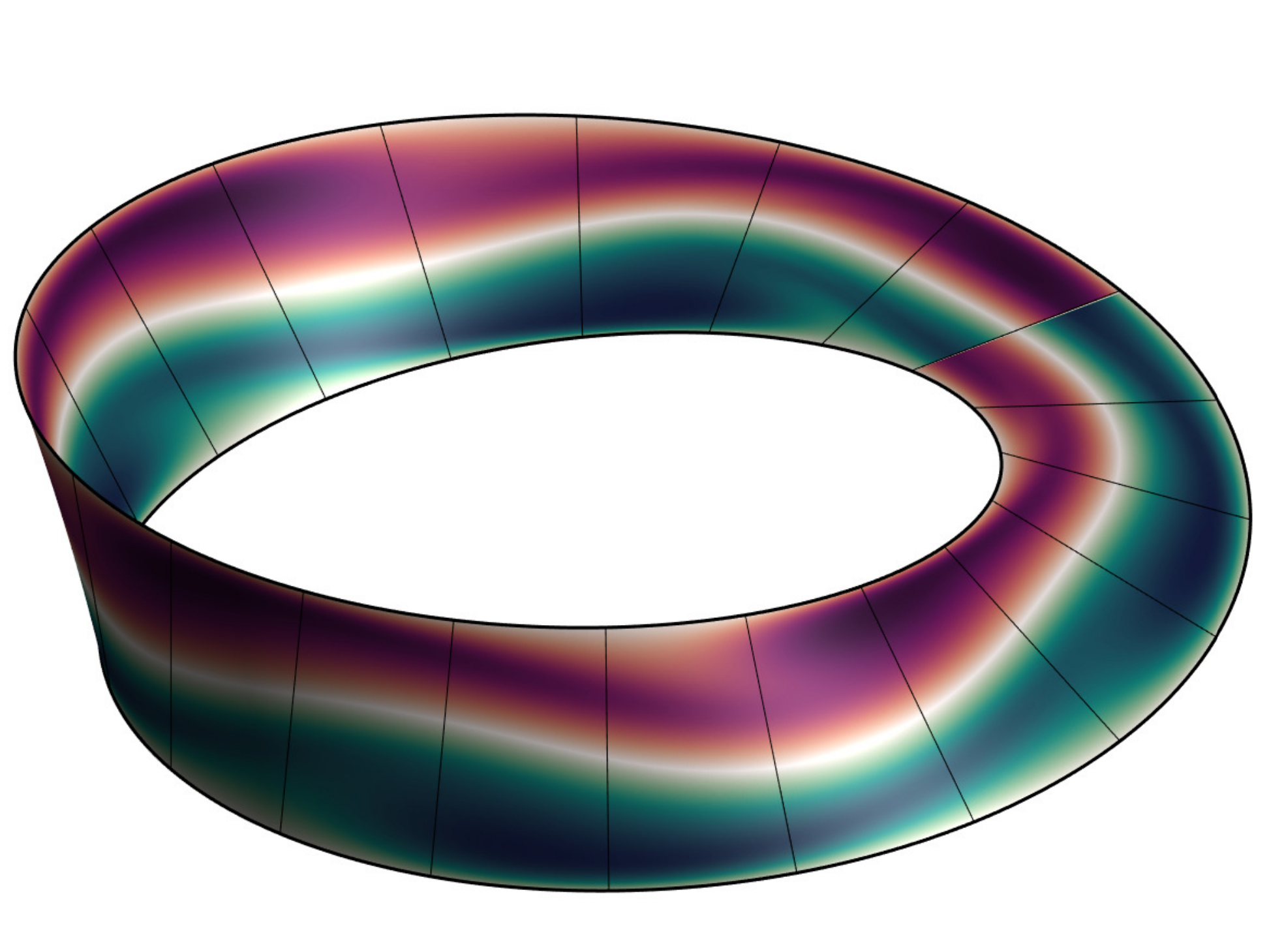} \hfill
\includegraphics[height=3.2cm]{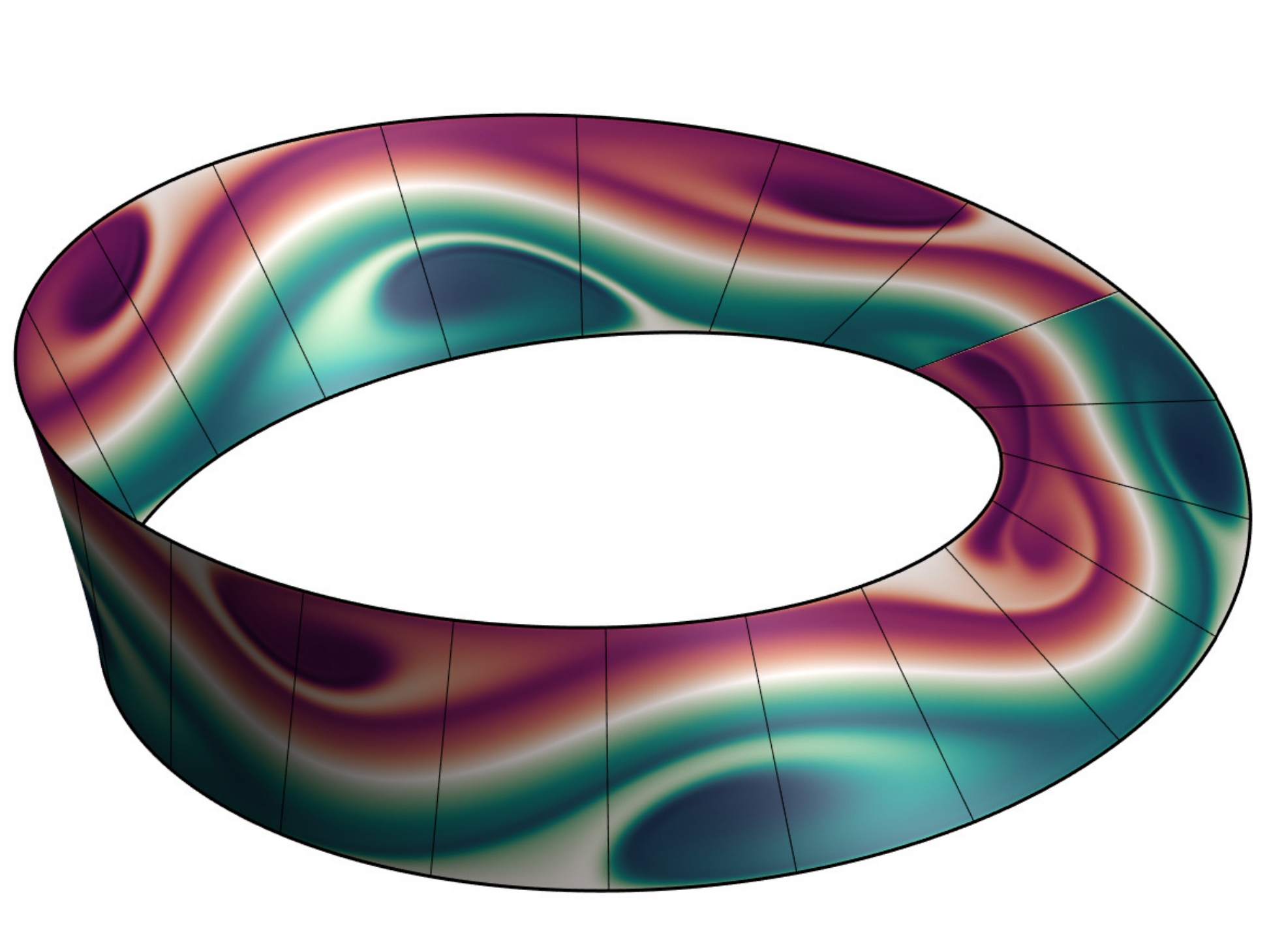} \hfill
 \includegraphics[height=3.2cm]{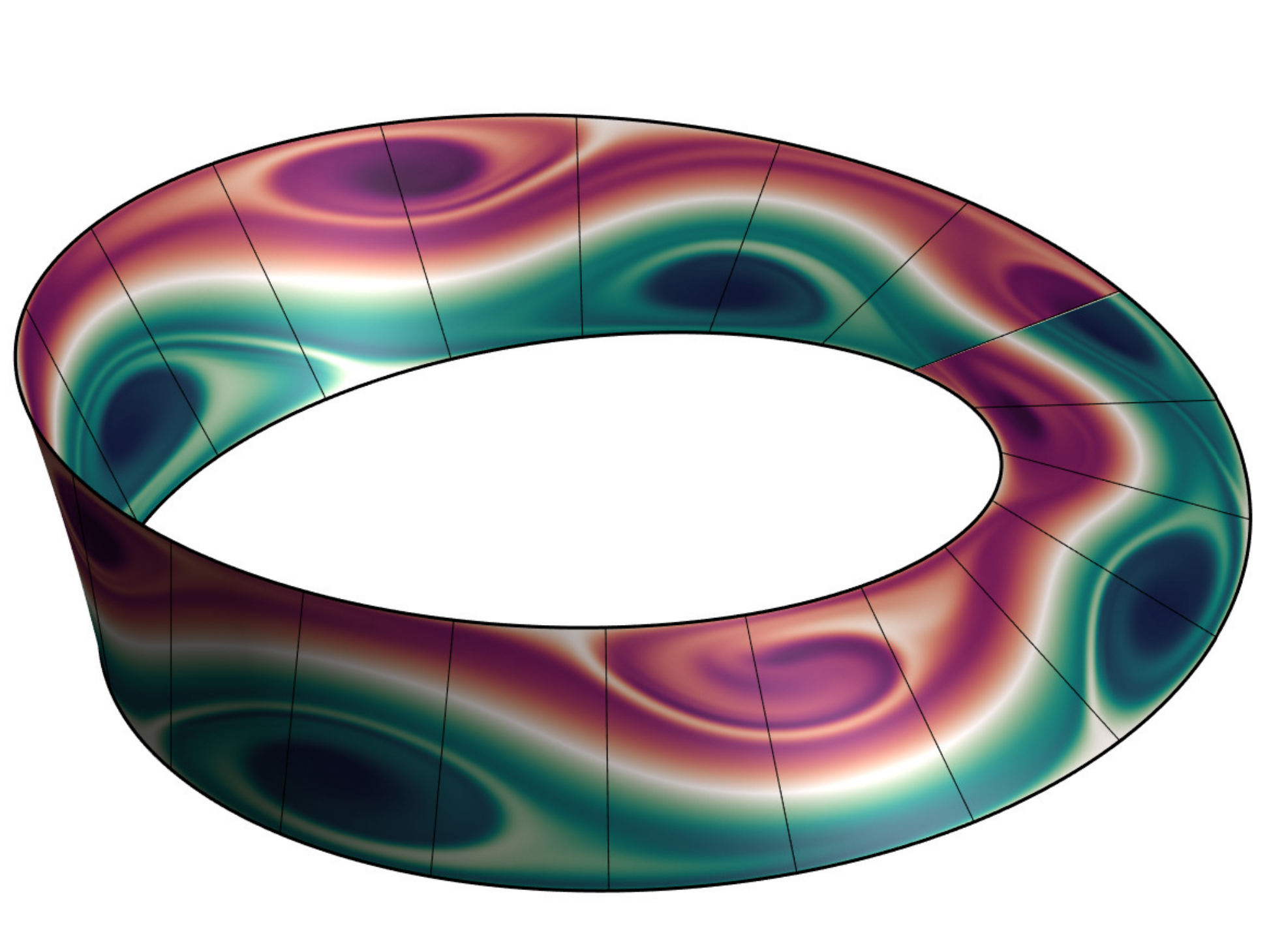}
 
\includegraphics[height=3.2cm]{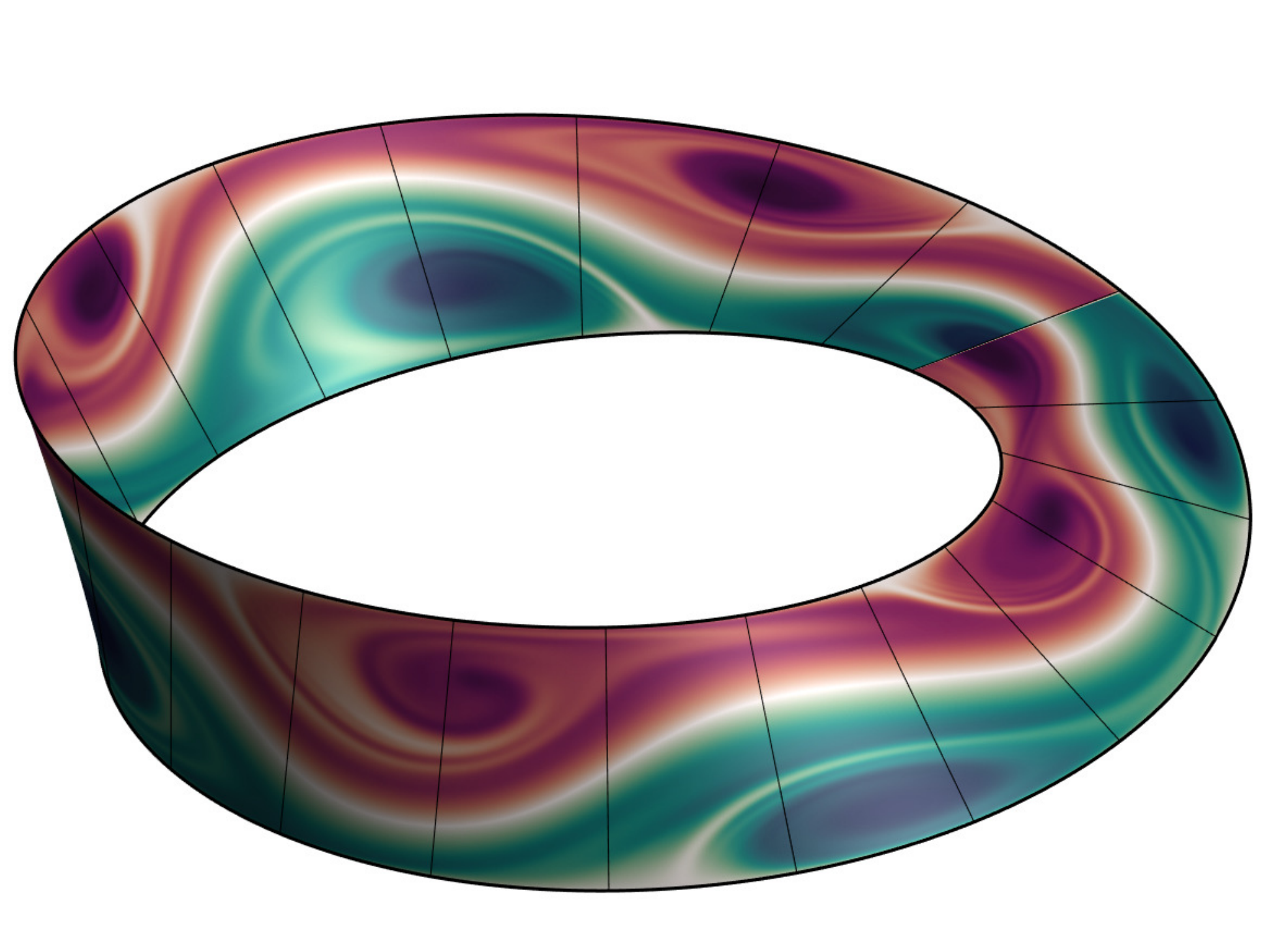} \hfill
\includegraphics[height=3.2cm]{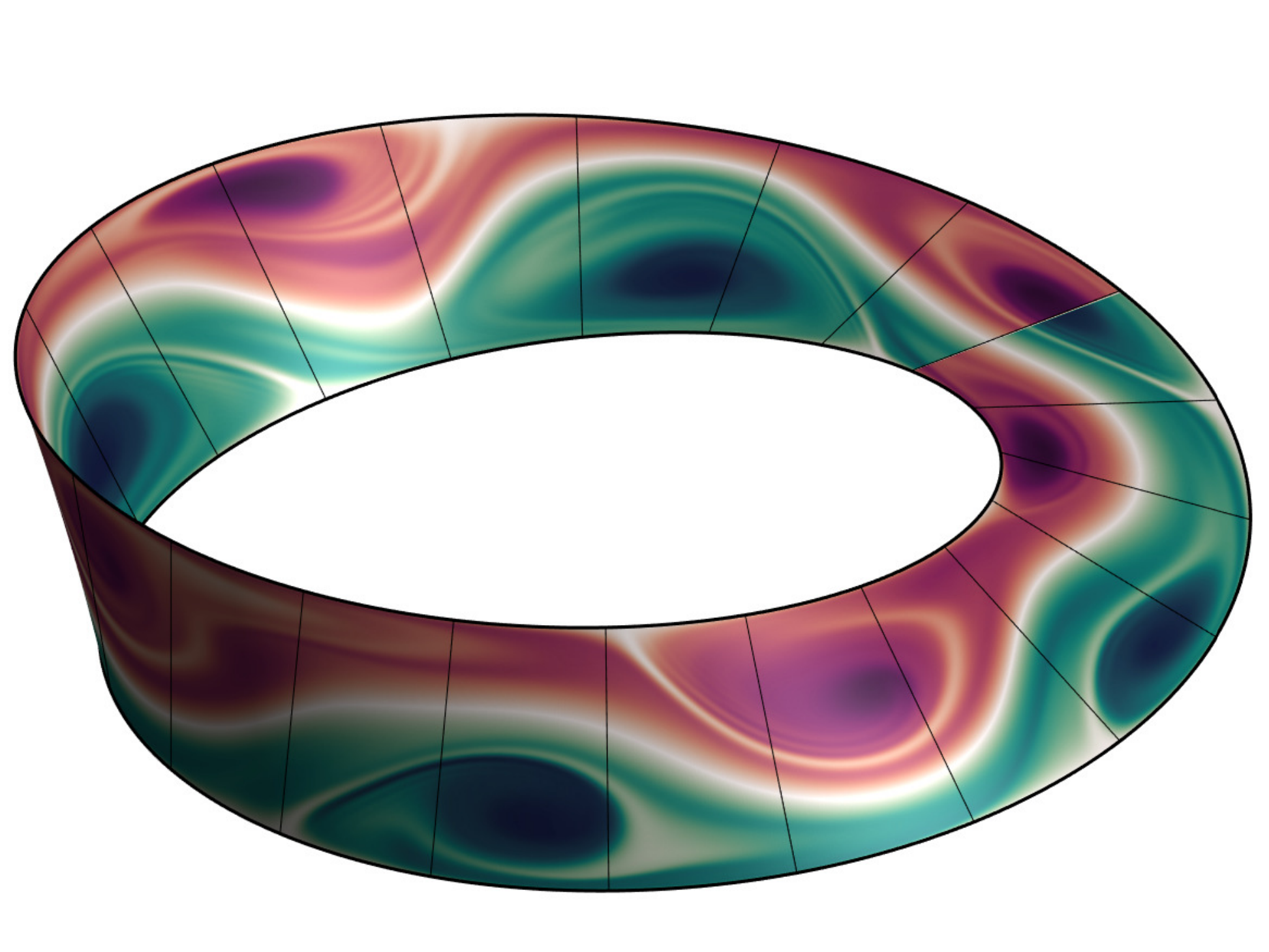} \hfill
 \includegraphics[height=3.2cm]{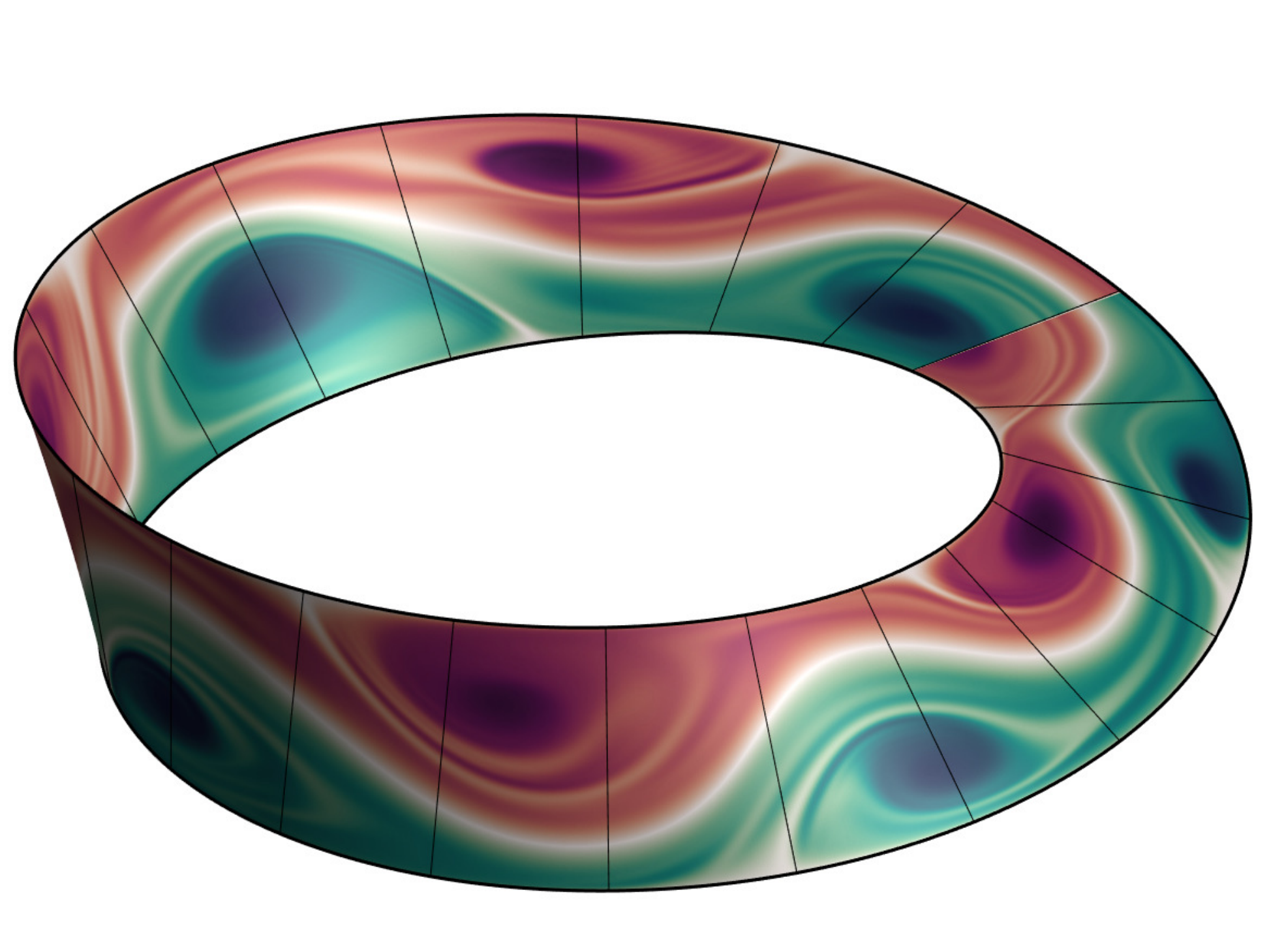} 
 \caption{Shear instability on a M\"obius strip: the {(pseudo-scalar) vorticity density} $\omega$ is shown at six successive time intervals, starting with a perturbation of the field \eqn{shear}
  (see also \href{http://www.maths.ed.ac.uk/~vanneste/mobius/vorticityMobiusShear.mp4}{movie~2}).}
 \label{fig:shear}
\end{center}
\end{figure}

We next examine a version of shear instability by simulating the evolution of the initial vorticity {density}
\beq
\omega=\zeta (a^2-\zeta^2).
\lab{shear}
\eeq
This is not a stationary solution of the Euler equations because the associated streamfunction depends on $\theta$ through the metric terms that appear in the Poisson equation \eqn{laplace}. Its evolution is nonetheless very slow compared to that which occurs when a small $\theta$-dependent perturbation is added -- we use a perturbation of the form $\omega' \propto (a^2-\zeta^2) \sin(8 \theta)$. The rapid evolution in the presence of a perturbation can be interpreted as resulting from shear instability. Note the relative complexity of \eqn{shear} compared with the quadratic vorticity profile  that is often employed to demonstrate shear instability in planar flows (a quadratic vorticity has a gradient vanishing at a single point, the inflection point of the velocity profile). This is imposed by the topological constraint that 2-forms, here $d \nu = \omega dx \wedge dy$, must vanish somewhere on non-orientable surfaces. Figure \ref{fig:shear} shows a series of snapshots of the pseudo-scalar vorticity {density} $\omega$ at equal time intervals. For distributed vorticity distributions, this field is easier to visualise than the vector field $\omega \bs{N}$ shown in figure \ref{fig:edge}, but it has the drawback of an abrupt sign change across the seam $\theta= 0$, visible near the top right in the representation of the M\"obius strip in figure \ref{fig:shear}. We emphasise that this discontinuity is artificial and that the physical fields $\nu$ and $d \nu$ (or equivalently $\omega \bs{N}$) {and the direction of rotation of the fluid} are continuous. The snapshots of figure \ref{fig:shear} and the accompanying \href{http://www.maths.ed.ac.uk/~vanneste/mobius/vorticityMobiusShear.mp4}{movie 2} show that shear instability on the M\"obius strip evolves very much as in planar channel, with the amplification of the initial wavy perturbation of the constant-vorticity lines leading to the formation of rows of counter-rotating vortices.

\begin{figure}
\begin{center}
\includegraphics[height=3.2cm]{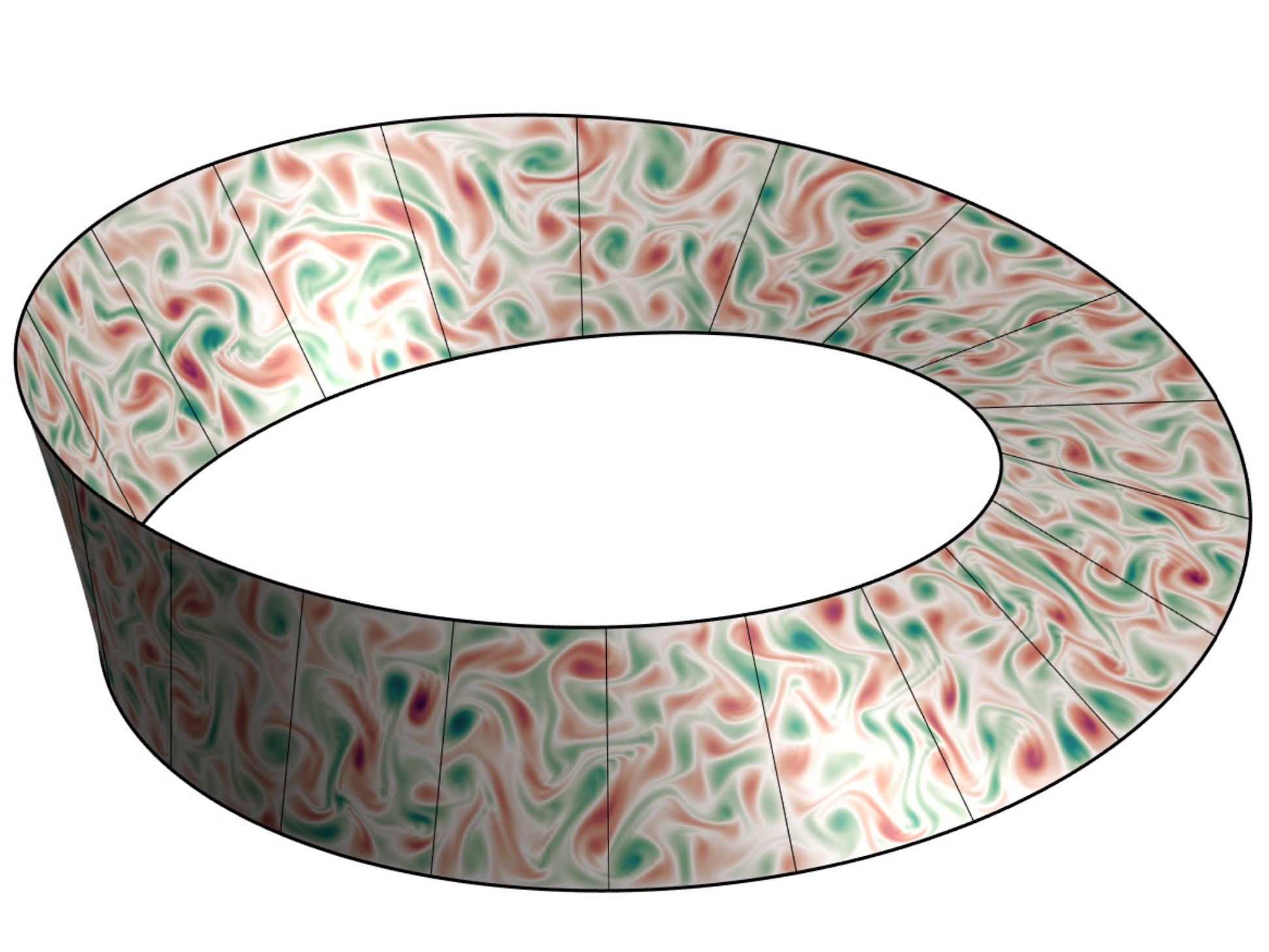} \hfill
\includegraphics[height=3.2cm]{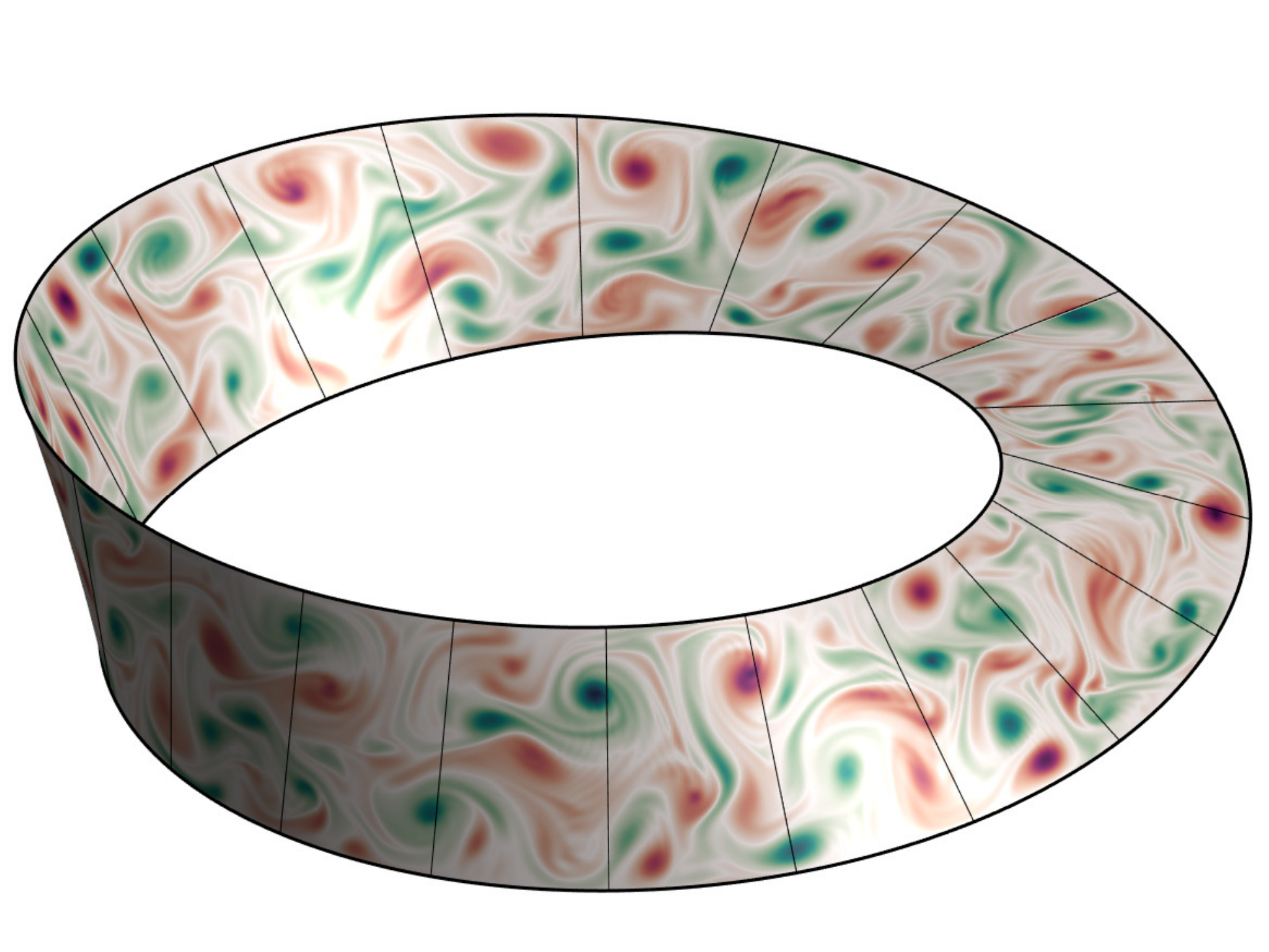} \hfill
 \includegraphics[height=3.2cm]{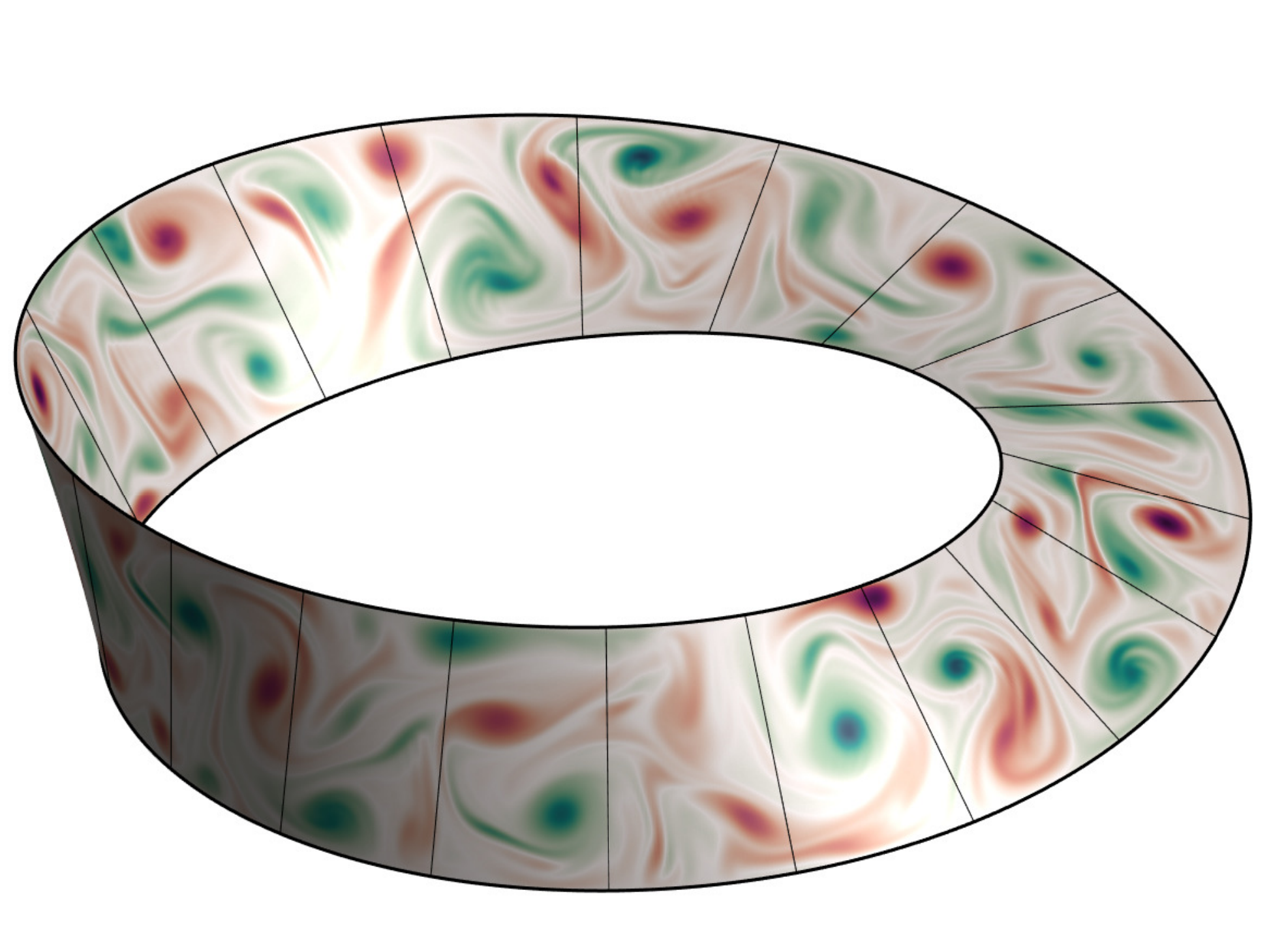}
 
\includegraphics[height=3.2cm]{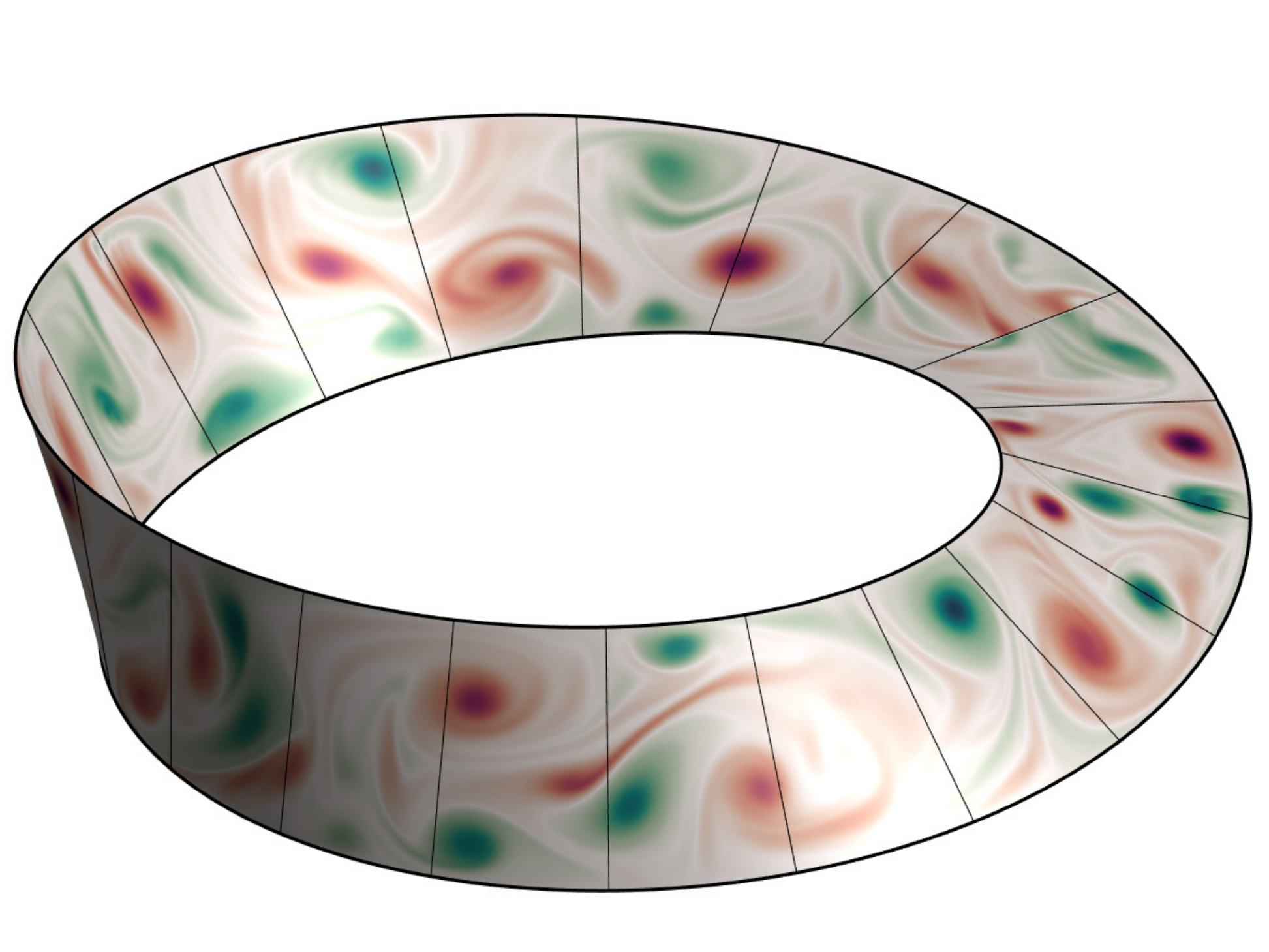} \hfill
\includegraphics[height=3.2cm]{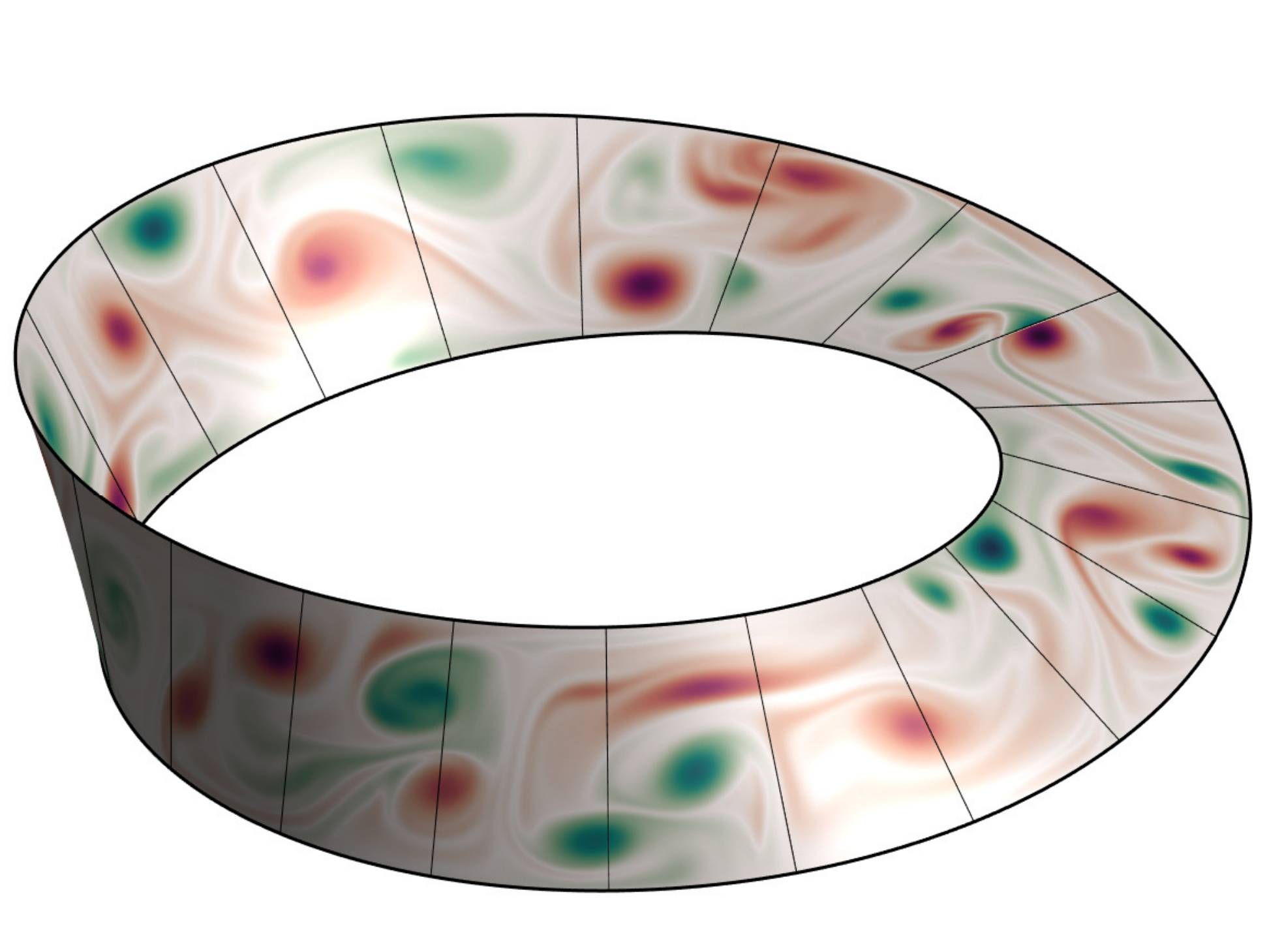} \hfill
 \includegraphics[height=3.2cm]{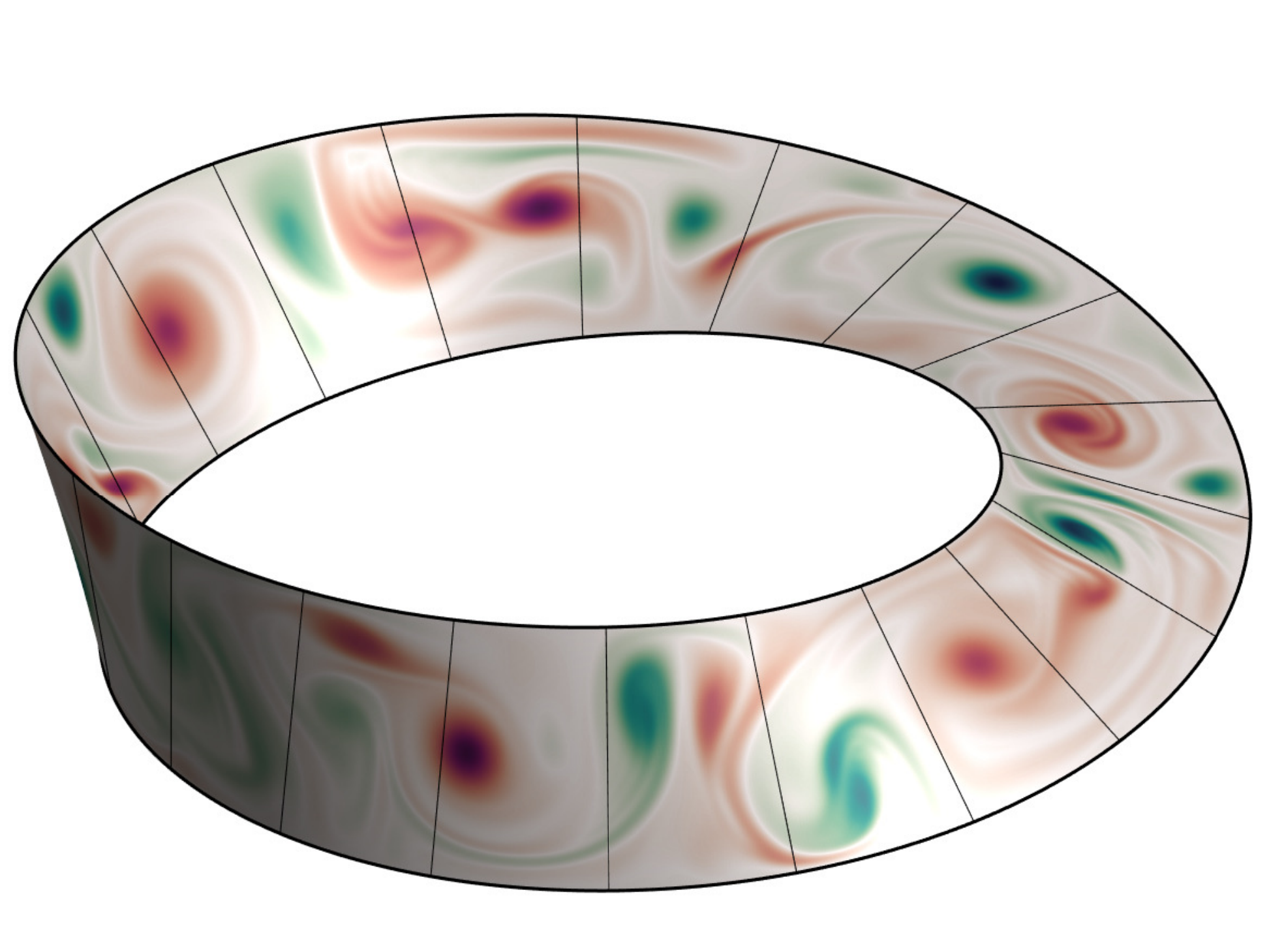} 
 \caption{Decaying turbulence on a M\"obius strip: the {(pseudo-scalar) vorticity density} $\omega$ is shown at six successive equal time intervals (see also \href{http://www.maths.ed.ac.uk/~vanneste/mobius/vorticityMobiusTurbulence.mp4}{movie 3}).}
 \label{fig:turbulence}
\end{center}
\end{figure}

Our final simulation, with results  displayed in figure \ref{fig:turbulence} and  \href{http://www.maths.ed.ac.uk/~vanneste/mobius/vorticityMobiusTurbulence.mp4}{movie 3}, illustrates the dynamics of decaying turbulence on the M\"obius strip. We build a small-scale, random initial {vorticity} density $\omega_0(\zeta,\theta)${, enforcing its pseudo-scalar property} by first constructing a random function $f(\zeta,\theta)$ that is $4a$-periodic in $\zeta$ and $2\pi$-periodic in $\theta$ using a standard Fourier series method. The initial vorticity {density} is then taken as
\beq
\omega_0(\zeta,\theta) = f(\zeta,\theta)-f(2a-\zeta,\theta)+f(2a+\zeta,\theta+\pi)-f(-\zeta,\theta+\pi).
\eeq
The periodicity of $f$ ensures that $\omega_0$ vanishes at the boundary and satisfies condition \eqn{pseudo-scalar} at the seam $\theta=\pi$, as required for a pseudo-scalar. The choice of peak wavenumbers $8\pi/a$ and $80\pi$ for $f$ leads to the reasonably isotropic, small-scale $\omega_0$ shown in the first panel of figure \ref{fig:turbulence}. In addition to an initial vorticity {density}, we also impose an non-zero (weak) initial circulation which leads to an overall drift of the vortices best seen in \href{http://www.maths.ed.ac.uk/~vanneste/mobius/vorticityMobiusTurbulence.mp4}{movie 3}.
The evolution shown  in figure \ref{fig:turbulence} and  \href{http://www.maths.ed.ac.uk/~vanneste/mobius/vorticityMobiusTurbulence.mp4}{movie 3} is characteristic of an inverse energy cascade, with vortex mergers that are local and hence unaffected by the topology of the strip. Over time scales much larger than those explored in our simulation, the scale of the vorticity field may become such that the evolution is more strongly influenced by the topology of $\mathcal{M}$. 

\section{Summary} \label{sec:final}

This paper examines how the Euler equations governing the dynamics of two-dimensional inviscid fluids can be formulated on a non-orientable surface, using a M\"obius strip embedded in $\mathbb{R}^3$ as a concrete example. The velocity vector field $\bu$, the corresponding momentum 1-form $\nu$ and the pressure field $p$ are all geometrically intrinsic objects, which are independent of the choice of coordinate basis regardless of whether the surface is orientable or not. Pseudo-fields, which depend on coordinates through the orientation of the basis, only appear when the Euler equations are simplified to their vorticity--streamfunction formulation. While the vorticity $d \nu$ is a proper 2-form, the more convenient $\omega$ is a pseudo-scalar because its definition via $d \nu = \omega \mu$ involves the area pseudo-2-form $\mu$. Similarly, the streamfunction $\psi$ is defined via the relation $d \psi = - \bu \ip \mu$ and is a pseudo-scalar. With our choice of coordinates $(\zeta,\theta)$, the difference between pseudo-scalars and scalars is simply that the former satisfy the twist  condition \eqn{bc1}  across $\theta = 0 \mod \pi$ instead of continuity. In the vorticity--streamfunction formulation of the Euler equations, the  condition that $\psi$ depends on time only along  each connected component of the boundary  must be supplemented by additional conditions tracing back to the original momentum formulation. On the M\"obius strip, the boundary is simply connected and the single function of time $C(t)$ that needs to be determined is found straighforwardly from conservation of the circulation along the single boundary. The main dynamical effect of  non-orientability is that the sign of $\omega$ can be changed by transport: a vortex patch rotating clockwise transported once along the strip returns rotating counterclockwise, as demonstrated in the numerical experiment of figure \ref{fig:edge}. This is why odd functions of the vorticity {density} are not conserved in an integral sense. Loosely speaking, the number of integral conservation laws on the M\"obius strip is half what it is on an orientable surface. It would be interesting to examine the consequences of this for the long-time dynamics of turbulence, e.g.\ as described by statistical mechanics. It would also be interesting to study purely non-dissipative dynamics on a M\"obius strip by considering point vortices or contour dynamics.  

\bigskip

\noindent
\textbf{Acknowledgments.} JV thanks Stefan Llewellyn Smith for pointing out useful references. 
\medskip

\noindent
\textbf{Declaration of interests.} The author reports no conflict of interest.

\appendix

\section{Momentum equation} \label{app:momentum}

We derive the momentum equation in coordinates $(\zeta,\theta)$. The derivation illustrates the convenience of working directly with forms rather than their components, as advocated by \citet{fran04}. Starting from the coordinate-free momentum equation \eqn{momentum} we compute
\begin{align}
(\partial_t + \lie_{\bu}) \nu &= (\partial_t + \lie_{\bu}) (u^\zeta \, d \zeta + |g| u^\theta \, d \theta) \nonumber \\
&= (\partial_t + \lie_{\bu}) (u^\zeta) \, d \zeta  + u^\zeta \, d (\lie_{\bu} \zeta) +  (\partial_t + \lie_{\bu}) (|g| u^\theta) \, d \theta + |g| u^\theta \, d(\lie_{\bu} \theta) \nonumber \\
&= (\partial_t + \lie_{\bu}) (u^\zeta) \, d \zeta  + u^\zeta \, d u^\zeta +  (\partial_t + \lie_{\bu}) (|g| u^\theta) \, d \theta + |g| u^\theta \, d u^\theta
\end{align}
and note that
\beq
d (p - \tfrac{1}{2} |\bu|^2) = d (p - \tfrac{1}{2} (u^\zeta)^2  - \tfrac{1}{2} |g| (u^\theta)^2) = d p - u^\zeta \, d u^\zeta - |g| u^\theta \, d u^\theta - \tfrac{1}{2} (u^\theta)^2 (\partial_\zeta |g| \, d \zeta + \partial_\theta |g| \, d \theta)
\eeq
to find the two components
\begin{subequations}
\lab{euler3}
\begin{align}
(\partial_t + \lie_{\bu}) u^\zeta  - \tfrac{1}{2} (u^\theta)^2 \partial_\zeta |g| &= - \partial_\zeta p,  \\
(\partial_t + \lie_{\bu})(|g| u^\theta)  - \tfrac{1}{2} (u^\theta)^2 \partial_\theta |g| &= - \partial_\theta p,
\end{align}
\end{subequations}
where $\lie_{\bu} = u^\zeta \partial_\zeta + u^\theta \partial_\theta$ since it acts on scalars.

\section{Geometric derivation} \label{app:hodge}

The vorticity--streamfunction formulation can be derived without resorting to coordinates. Following \citet{kimu99} and \cite{raga-barr}, this is best carried out using the Hodge * operator which, in two dimensions, is determined by the following:
\beq
*1 = \mu, \quad * \alpha = \alpha^\sharp \ip \mu \inter{and} *\mu=1, 
\eeq
where $\alpha$ is a 1-form and $\alpha^\sharp$ its dual vector, with $\alpha=g(\alpha^\sharp,\cdot)$. In coordinates, we have
\beq
* (\alpha_\zeta \, d \zeta + \alpha_\theta \, d \theta) = |g|^{1/2} \alpha_\zeta d \theta - |g|^{-1/2} \alpha_\theta d \zeta. 
\lab{staralpha}
\eeq
Noting that $** \alpha = - \alpha$ \citep[e.g.][]{fran04}, we can rewrite \eqn{psiu} and \eqn{laplace} as
\beq
\nu = *d \psi \inter{and} \omega = * d \nu =  * d * d \psi,
\eeq
which identifies the Laplacian acting on (pseudo-)scalars as $\Laplace = * d * d$. 

This form of the Laplacian is also useful to compute the change in the circulation along the boundary that is introduced by  dissipation. For simplicity, we take the our dissipative model to be
\beq
(\partial_t + \lie_u) \nu = \varepsilon \Laplace \nu,
\eeq
where $\Laplace = d*d* + *d*d$ is the Laplace--de Rham operator. We emphasise that this is not the standard (Navier--Stokes) viscous dissipation which, instead, involves a different Laplacian defined via the divergence of the stress tensor \citep{gilb-et-al,gilb-v21}. Integrating over the (closed) boundary $\partial \mathcal{M}$ and noting that the outer exact differential in $\Laplace$ integrates to zero gives
\beq
\dt{\Gamma}{t} = \varepsilon  \int_{\partial \mathcal{M}} *d * d \nu = \varepsilon  \int_{\partial \mathcal{M}} *d \omega,
\eeq 
leading to the coordinate expression
\beq
\dt{\Gamma}{t} = \varepsilon \int_0^\pi \left( |g|^{1/2}(d,\theta) \partial_\zeta \omega(d,\theta,t) + |g|^{1/2}(-d,\theta) \partial_\zeta \omega(-d,\theta,t) \right) \, d \theta
\lab{circchange}
\eeq
on using \eqn{staralpha} and that $\omega=0$ on $\partial \mathcal{M}$.

\bibliographystyle{jfm}
\bibliography{mybib}


 
\end{document}